\documentclass[twocolumn,showpacs,superscriptaddress,preprintnumbers,amsmath,amssymb,pre]{revtex4}
\usepackage{graphicx}
\usepackage{dcolumn}
\usepackage{bm}
\usepackage{amsmath}
\usepackage{epstopdf}
\usepackage{verbatim}
\newcommand{\gammadot}{\dot{\gamma}}

\begin{document}

\title{Rheology of Weakly Vibrated Granular Media}

\author{Geert H. Wortel}
\affiliation{Kamerlingh Onnes Lab, Universiteit Leiden, Postbus
9504, 2300 RA Leiden, The Netherlands}

\author{Joshua A. Dijksman}
\affiliation{Kamerlingh Onnes Lab, Universiteit Leiden, Postbus
9504, 2300 RA Leiden, The Netherlands}\affiliation{Dept. of Physics, Duke University, Science Drive, Durham NC 27708-0305,USA}

\author{Martin van Hecke}
\affiliation{Kamerlingh Onnes Lab, Universiteit Leiden, Postbus
9504, 2300 RA Leiden, The Netherlands}

\date{\today}

\begin{abstract}
We probe the rheology of weakly vibrated granular flows as function of flow rate,  vibration strength and pressure by performing experiments in a vertically vibrated split-bottom shear cell. For slow flows, we establish the existence of a novel vibration dominated granular flow regime, where the driving stresses smoothly vanish as the driving rate is diminished. We distinguish three qualitatively different vibration dominated rheologies, most strikingly a regime where the shear stresses no longer are proportional to the pressure.
\end{abstract}

\maketitle

Granular media are collections of macroscopic, athermal grains which interact through dissipative, frictional contact forces. In the presence of gravity and in the absence of additional external forces, they jam in metastable configurations; however, external forcing can easily lead to yielding and flow  \cite{1995_book_duran,1996_revmodphys_jaeger,2004_epje_gdrmidi,2008_annurevfluid_forterre,2010_prl_nichol,
2007_pre_sanchez, 2006_pre_rubin,1989_prl_jaeger, 2009_jrheol_marchal, 2009_epl_janda}.
The best known scenario that leads to granular flow is by exerting shear stresses that exceed the yield stress, as when tilting a quiescent layer of sand sufficiently far away from the horizontal \cite{1995_book_duran,1996_revmodphys_jaeger}. To understand such dense granular flows, it is becoming
increasingly clear that both stress and mechanical agitations play a crucial role. Indeed, a given stress can give rise to a wide range of flow rates depending on the magnitude of these agitations \cite{2011_prl_reddy,2010_prl_nichol, 2013_elie_arxiv, andreotti_arxiv}.  Moreover, agitations make granular media lose their rigidity, although in the absence of shear stresses this does not need to cause flow  \cite{2010_prl_nichol, 2011_prl_reddy, 2012_pre_nichol, 2013_elie_arxiv}. We note here that the idea that both the stress and the amount of agitations determine the flow rate lies at the basis of numerous models for slowly flowing disordered materials \cite{1998_pre_falk,1997_prl_sollich,2012_pre_krimer,2008_nature_goyon}.

\begin{figure}[t!]
	\begin{center}
    \includegraphics[width=5cm]{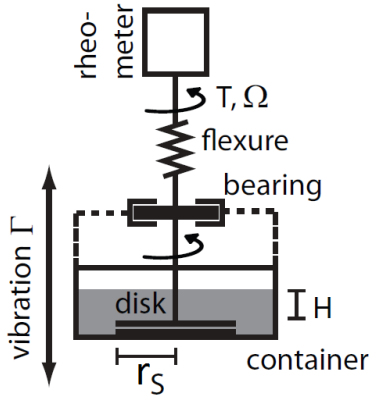}\caption{Sketch of the vibrated split-bottom setup in which the rotation of a disk of radius $r_s$ is used to probe the rheology of agitated granular media. The crucial experimental parameters are the filling height, $H$, vibration amplitude, $\Gamma$, the torque, $T$, and the rotation rate, $\Omega$.}\label{fig:intro}
\end{center}
\end{figure}

In a granular context, such agitations may be provided by external vibrations. In a classic experiment, the slope of a granular pile was found to relax under vertical vibrations~\cite{1989_prl_jaeger}, and similarly, horizontal vibrations have been
used to induce flow on inclined planes~\cite{2007_pre_sanchez, 2006_pre_rubin}.
Piezo transducers inside the medium have been used to inject tiny rearrangements or force fluctuations, either locally~\cite{2009_epl_janda, 2009_epje_caballero} or along a complete boundary~\cite{2005_jphyscm_caballero}.  Shear induced agitations similarly have induced microscopic rearrangements~\cite{2012_epl_coulais}.

Naturally, flow itself also induces mechanical agitations~\cite{2002_prl_longhi}. For example, for shear banded flows \cite{schall}, the observation of particle rearrangements and fluidization far away from the flowing region suggest that the effect of agitations can be carried far through the material
\cite{2010_prl_nichol,2008_jstatmech_crassous, 2011_prl_reddy, 2012_prl_amon}.
Agitations thus form a crucial ingredient for non-local extensions of models for the rheology of granular flows~\cite{2008_annurevfluid_forterre, 2008_pre_jagla, 2007_pre_torok, 2006_pre_depken}, and may explain, among other things, the large extension of shear bands in split-bottom granular flows \cite{2012_prl_kamrin,PNAS_kamrin}.

Recently, we have explored how weak vibrations influence the rheology of dry granular media~\cite{2011_prl_dijksman} by performing experiments in a vibrated split-bottom cell, as shown in Fig.~\ref{fig:intro}. We found that weak external vibrations suppress the yield stress of the material and strongly influence the rheology of slow granular flows.

In this paper, we reveal the intriguing rheology of weakly vibrated granular media in much more detail. We find that we can distinguish a variety of qualitatively different flow regimes. First, for large flow rates, inertial effects dominate, and the effect of vibrations is small. Second, for slower flow rates, we cross over to a regime similar to the well-known quasi static flows that have been studied at length in the absence of vibrations \cite{1980_powtech_nedderman, 2004_prl_fenistein, 2006_prl_fenistein, 2010_sm_dijksman, 2011_prl_dijksman}. Third, for slower flows, we enter a regime where the vibrations lead to completely new rheological behavior.

The focus of this paper is on these {\em vibration dominated flows}. By probing the equilibration times of the stresses and the variation of the steady state stresses with filling height, we find evidence for three qualitatively different regimes. For slow enough flows, vibration effects increasingly
dominate the physics, leading to compaction of the material for weak vibrations, and to fluidization of the material for vibrational accelerations approaching gravity. Most strikingly, in the latter regime, we see a breakdown of the proportionality of shear stresses and pressure, a highly unusual phenomenon in granular flows.

We also study how the rheology of weakly vibrated granular media behaves in stress controlled flow experiments. We find that equilibration times can be dramatically longer than in rate controlled experiments, akin to what has been observed in several soft materials that exhibit a so-called viscosity bifurcation \cite{coussot_pre_2002,bonn_epl_2009}, and leading to Andrade creep like phenomena \cite{andrade_1914}, providing an interesting analogy with thermal, disordered flows~\cite{2004_jphyscondmat_petekidis, ballauff_prl_2012}.

The outline of this paper is as follows. In Section~\ref{sec:setup} we describe the details of our experimental setup and the measurement protocols used in the current and previous study~\cite{2011_prl_dijksman}. In Section~\ref{sec:pheno} we describe the main phenomenology of a complete set of experiments probing $T(\Gamma,\Omega,H)$. In Section ~\ref{sec:unger} we introduce the canonical perspectives on granular rheology, including a model for the stresses in split-bottom flows by Unger {\em et al.} \cite{2004_prl_unger}. In Section~\ref{sec:Ggt0} we use this model to extract effective friction coefficients from our data, as well as exploring the quality of the fit between the data and this model. In Section~\ref{sec:fluid} we provide strong evidence for the existence of a pressure independent flow regime though measurements of the flow structure. Additionally, we compare steady state results obtained at constant $\Omega$ driving with a constant stress driving mode in Section~\ref{sec:constantT}.

\section{Setup and Protocol} \label{sec:setup}
In this section we briefly introduce the main parts of our experimental setup (for details, see Appendix \ref{app:setup}), discuss our measurement protocols, and show how we ensure that we measure steady values for the rheology.

\subsection{Setup}
We probe the rheology of weakly vibrated granular flows in a modified split-bottom cell, as shown in Fig.~1. The acrylic container has an inner radius of 7~cm. We mount a hollow smooth cylinder of 6~mm height and 4~cm radius on the bottom of the container. The rotating disk (radius $r_s$ of 4~cm and thickness 5~mm) that drives the granular flow is mounted just above the cylinder. The gap between the container and the disk is about 0.3~mm so no particles can get underneath the disk. To ensure a no-slip boundary condition, the top surface of the disk is made rough by gluing glass particles with diameter of 2~mm to it.

The container is filled with particles (black soda-lime glass beads, Sigmund Lindner 4504-007-L), a polydisperse mixture with a diameter between 1 and 1.3~mm, and a bulk density $\rho$ of 1.7$\times 10^3$~kg/m$^3$, up to a filling height $H$. To ensure good reproducibility, we use to total mass of the particles to control $H$.

All experiments are carried out under ambient temperature, pressure and relative humidity. We have verified that our experiments are insensitive to relative humidities ranging from 6 and 55\%. After several months of use, the black coating of the particles visibly deteriorates, and the rheological behavior becomes more sensitive to humidity. We therefore renew our particles on a trimonthly basis, and found that our experiments reproduce well over the course of several years.

In the absence of vibrations, the phenomenology of the flow is determined by the dimensionless filling height $h\equiv H/r_s$ \cite{2010_sm_dijksman}. In our experiments, we stay in the low filling height regime ($h < 0.6$), where the shear bands are mainly vertical, and all grains above the disk co-rotate along with it (trumpet-flow). We have found that in all but one flow regimes, the flow profiles observed at the free surface are insensitive to the magnitude of vibrations.  For the exception -- which is for slow flow and strong vibrations -- see Section \ref{sec:fluid}.

To drive the rotation of the bottom disk and shear the granular media we use a rheometer (Anton Paar DSR 301), which can be used both in stress control (imposing a torque $T$ and measuring the resulting rotation rate $\Omega$) or in rate control (impose $\Omega$, measure $T$). We shake the system with a sinusoidal oscillation $Asin(2\pi f t)$, with a fixed $f$ of 63~Hz, using an electromagnetic shaker (VTS systems VG100). The amount of vibrations is characterized by the dimensionless parameter $\Gamma$=$A(2\pi f)^2/g$, where g is the gravitational acceleration. We image the surface of the system  using a mirror and a Foculus FO114B camera, allowing us to extract the surface flow by particle image velocimetry.

\begin{figure}[t!]
    \begin{center}
		\includegraphics[width=\columnwidth]{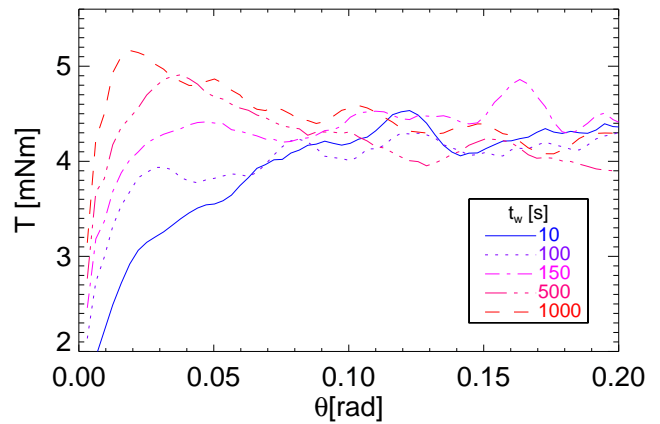}
        \caption{(Color online) The torque as function of deflection angle $\theta$ for $\Omega$=10$^{-4}$~rps and $\Gamma=0.7$ but different waiting time $t_w$ between the preshear and the actual measurement. }\label{fig:waiting}
    \end{center}
\end{figure}
\subsection{Protocol}
Our experiments focus on the rheological curves which relate the driving torque $T$ and the driving rate $\Omega$.
In section \ref{sec:constantT} we describe some experiments performed at constant torque, but our main focus is on experiments where we fix the driving rate in the range from 1 to 10$^{-4}$~rps, and probe the torque. We perform these experiments for a range of vibration amplitudes $\Gamma = 0, 0.2, 0.5, 0.7, 0.83, 0.95$ and $1$, and moreover
use  seven different filling heights ($h = 0.19, 0.25, 0.31, 0.38, 0.44, 0.50$ and $0.56$). Varying $h$ allows us to probe the role of the confining pressure for the rheology.

Each experiments starts with switching on the vibrations, after which we allow the shaker feedback loop 30~s to settle to the required value of $\Gamma$. We proceed by applying an amount of pre-shear to the granular material, in order to obtain similar starting conditions for each experiment. Unless noted otherwise, the protocol consists of the following steps: \textit{(i)} 2~s of 1~rps rotation clockwise; \textit{(ii)} 4~s of 1~rps rotation counter-clockwise; \textit{(iii)} 2~s of 1~rps rotation clockwise. \textit{(iv)} 5~s without imposed stress or shear. \textit{(v)} start of actual measurement. The rotation in the experiments is in the clockwise direction to minimize anisotropy effects \cite{wortel_aniso}.

Vibrations lead to compaction of granular media, although this process is very slow for $\Gamma < 1$ \citep{2005_jphyscm_caballero}, whereas flow typically leads to dilatation \cite{1885_philmag_reynolds}. Additionally, anisotropy in the fabric of the granular media needs a finite amount of strain to build up, but may be relaxed by vibrations \cite{wortel_aniso,2013_elie_arxiv}. For both density and anisotropy,
vibrations and flow are in competition, and as we are interested in steady state flow properties, we need to ask: what is the minimum time or strain necessary to get into a steady state flow regime?

We have probed the relaxation of our flows to a steady state by starting the flow from a denser or less densely packed state as follows: Before each experiment, we perform pre-shear as described above. After pre-shear, during stage \textit{(iv)}, we stop the shear and vibrate the material for a waiting period $t_w$, during which the granular packing density increases by compaction, and then start the actual measurements. By varying the $t_w$, we thus manipulate the packing fraction at the start of the flow.  By measuring the torque as function of time we capture the evolution of the torque to its steady state value. As we expect this equilibration to be slowest for small $\Omega$ we perform this test at the smallest $\Omega$ (10$^{-4}$~rps) that we explore in our experiments.

\begin{figure}[t!]
    \begin{center}		
        \includegraphics[width=\columnwidth]{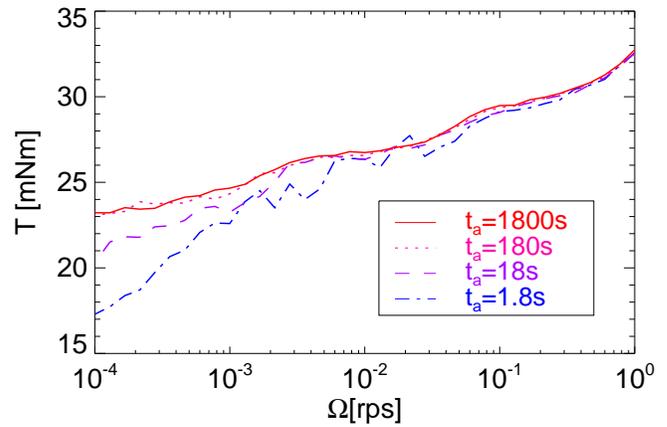}
        \caption{(Color online) Flow curves for different averaging times $t_a$. Below $\Omega \approx 0.5\times10^{-3}$, the torque increases with the waiting time. The measurements are for $\Gamma=0.6$ and $h$=0.56.}\label{fig:aging}
    \end{center}
\end{figure}

The results of this test are shown in  Fig.~\ref{fig:waiting}, where we plot $T$
as function of the total angle of rotation of the bottom disk $\theta$. This figure shows that for small $t_w$, $T$ grows monotonically before reaching steady state, whereas for large $t_w$, the torque  peaks at values larger than the steady state value. This is consistent with a simple picture where the longer the waiting time, the denser the grains are packed at the start of the experiment, and the larger the torque needed to cause flow.

For all waiting times, the torque reaches its steady state value for $\theta < 0.1$~rad, corresponding to a measurement time of 150~s at 10$^{-4}$~rps. We claim that this represents the longest equilibration time necessary to reach a steady state flow situation, as all our experiments are carried out for $\Omega \geq 10^{-4}$~rps. Moreover, in many experiments our data is acquired in a so called strain rate sweep, where the rotation rate is varied by a small amount so that equilibration will be faster. In all cases, an equilibration strain or time of $\theta > 0.1$~rad, or 150~s, will be sufficient to obtain steady state flow curves. We choose 180~s for all the experiments described in this paper.

To independently verify that equilibration times of 180~s are long enough, we perform a strain rate sweep at fixed $\Gamma=0.6$. We sweep the flow rate from fast to slow rates, and then compare flow curves obtained for different times $t_a$ per step, as shown in  Fig.~\ref{fig:aging}. While for small values of $\Omega$ and $t_a$, the torque shows a variation with $t_a$, for all the flow rates probed here we note that the data for $t_a=180$~s and $1800$~s are indistinguishable, showing that for $t_a \geq 180$~s the obtained values of $T$ are steady state values. As a final additional test, we have also inspected $T(t)$ to confirm we reach steady state (Appendix~\ref{proto}).

\section{Phenomenology} \label{sec:pheno}
\begin{figure*}[t!]
\begin{center}
\includegraphics[width=17cm]{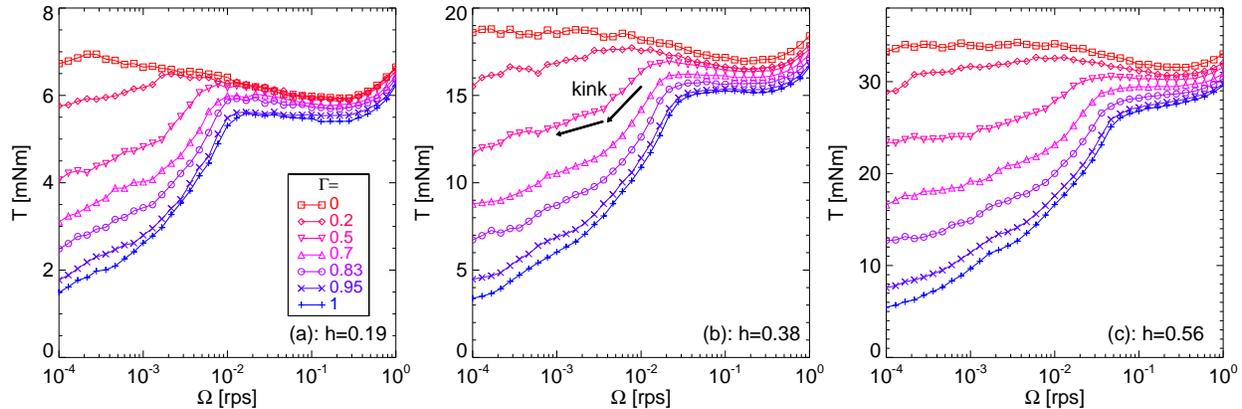}
\caption{(Color online) Selected flow curves for fixed filling height and varying $\Gamma$. In all cases, $T$ decreases monotonically with $\Gamma$ --- its dependence on $\Omega$ is more complex. The selected filling heights are $h= 0.19$ (a), $h= 0.38$ (b),  $h= 0.56$ (c).}\label{fig:overview_g}
\end{center}
\end{figure*}
\begin{figure}[b!]
    \begin{center}
				\includegraphics[width=\columnwidth]{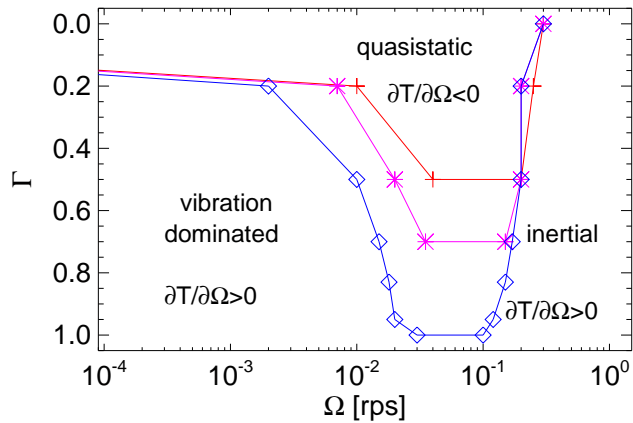}
        \caption{(Color online) The boundaries between the regions where the flow curves have a positive and a negative slope for $h$= 0.19 ($\diamond$), $h$= 0.38 ($\ast$) and $h$= 0.56 (+). For $\Gamma=0$ and $\Omega<0.3$~rps, the flow curve always has a negative slope. The region extends to $\Gamma>0$, and it extends to higher $\Gamma$ for lower $h$.}\label{fig:slope}
    \end{center}
\end{figure}

We now turn our attention to the rheological curves $T(\Omega)$. As shown in the $T(\Omega)$ curves in Fig.~\ref{fig:overview_g}, the flows in our system exhibit a rich rheology. There are two simple trends we see illustrated in these curves: increasing the filling height always increases $T$, whereas increasing the vibration strength always decreases $T$. The role of the flow rate is not as simple, with the torque often being a non monotonic function of the disk rate $\Omega$ ---  moreover, the details of the rheological curves depend on both the vibration strength $\Gamma$ and filling height $h$. We note here that the sign of $\partial T/ \partial\Omega$ has a crucial rheological implication: flows for which $\partial T/ \partial\Omega > 0$ can also be accessed in experiments where the torque is fixed, whereas flows for which $\partial T/ \partial\Omega < 0$ are {\em unstable} in torque controlled experiments. As we discussed in \cite{2011_prl_dijksman}, this range of unstable flows leads to hysteretic switching between two different flow regimes when the torque is varied, and is deeply connected to the yielding behavior of granular media observed for $\Gamma=0$.

Here we focus on rate controlled experiments, and as a first step in characterizing these curves, we plot the boundaries between the regions where $\partial T/ \partial\Omega$ is positive and negative for three values of $h$ in Fig.~\ref{fig:slope}. Roughly speaking, we can distinguish three regimes.

{\em Inertial flows ---} For $\Omega \gtrsim 0.3$~rps, $\partial T/ \partial\Omega > 0$; the flow curves show an increasing $T$ for increasing $\Omega$. This increase corresponds to the onset of the inertial regime \cite{pouliquen_I}. To estimate the inertial number $I=\dot{\gamma}d/ \sqrt{P/\rho}$ at $\Omega=0.3$~rps, we have to choose a characteristic pressure and strain rate scale, as both $\dot{\gamma}$ and $P$ vary throughout the system.
Taking $P$ as the hydrostatic pressure at $0.5H$, and $\dot{\gamma}$ corresponding to a shear band of 3 particles wide, we get $I=0.09$ for $h=0.38$ and $\Omega=0.3$~rps. Considering that the inertial regime typically starts at $I=0.1$~\cite{2006_nature_jop}, there is good agreement between the onset of increasing $T(\Omega)$ and the onset of the inertial regime. In the remainder of the paper we will focus on slower flows.

{\em Unstable flows ---} For intermediate flow rates, $T(\Omega)$ has a negative slope for small $\Gamma$ --- for $\Gamma \rightarrow 0$, this regime extends to arbitrarily small flow rates, although there the flow curves become essentially flat.
Despite the unstable character of the global rheology, and in contrast to unstable flows in e.g. micelles \cite{schall}, we do not see any changes in the shear bands as we move in and out of this unstable regime. As the variation of the stress with flow rate is not very large, this regime can also be referred to as {\em quasistatic}.

{\em Vibration dominated flows ---} Both the unstable/quasistatic and inertial regime have been studied in great detail already~\cite{1980_powtech_nedderman,2006_pre_depken,2007_epl_depken,2004_epje_gdrmidi,2006_nature_jop,2008_annurevfluid_forterre}, as they also arise in the absence of vibrations. Hence, in the remainder of the paper we will focus on the new vibration dominated regime of slow, stable flows that arises for $\Gamma >0$ and $\Omega < 10^{-2}$~rps.

One striking qualitative feature of this regime we already want to point out is the pronounced ''kink'' in the flow curves that can be seen for $10^{-2}$~rps$<\Omega<10^{-3}$~rps in Fig. \ref{fig:overview_g}. In Fig. \ref{fig:overview_g}b we explicitly mark such a kink. The kinks coincide with the flow rates where $t_a$ needs to be sufficiently large for $T$ to equilibrate (see Fig.~\ref{fig:aging}). We suggest that at sufficiently low $\Omega$, compaction effects become significant, leading to an increase of $T$ with time, and a ''kink'' in the flow curves.

\begin{figure*}[t!]
\begin{center}
\includegraphics[width=17cm]{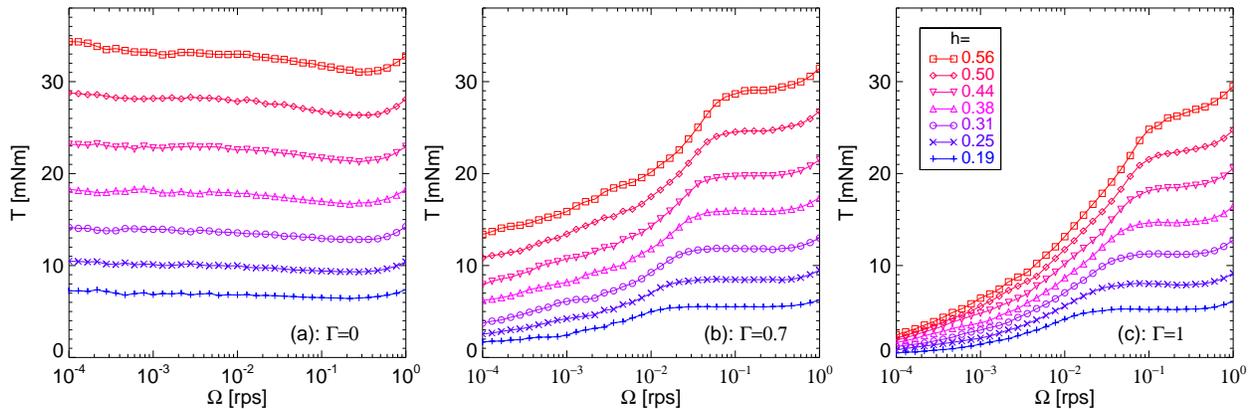}
\caption{(Color online) Selected flow curves for fixed $\Gamma$ and varying $h$. In all cases, $T$ grows monotonically with $h$. The selected values of $\Gamma$ are: (a) $\Gamma=0$, (b) $\Gamma=0.7$ and (c) $\Gamma=1$. }\label{fig:overview_h}
    \end{center}
\end{figure*}

\section{Vibration Dominated Flows} \label{granorl}
We will now turn our attention to the increase of $T$ with $h$, which allows us to probe the underlying mechanisms that govern the rheology of vibration dominated flows.
The canonical starting point of descriptions of {\em non-vibrated} slow granular flows is that the shear stresses $\tau$ are proportional to the pressure $P$ \cite{1980_powtech_nedderman,2006_pre_depken,2007_epl_depken}, and the ratio of $\tau$ and $P$ is an effective friction coefficient, $\mu$. For inertial flows, a description where $\mu$ becomes rate dependent (through the inertial number) has been shown to capture much of the phenomenology \cite{2004_epje_gdrmidi,2006_nature_jop,2008_annurevfluid_forterre}, and for slow, non-vibrated flows, this Mohr-Coulomb picture combined with a non-local rheology captures the essentials of steady, slow granular flow \cite{2012_prl_kamrin,PNAS_kamrin}.

By varying the filling height $h$, we can modify the pressure $P$ and probe its role for the rheology in the different regimes. Clear predictions for $T(h)$ exist from a well-studied rheological model for the driving torques in a split-bottom geometry \cite{2004_prl_unger}. In addition, this model provides clues to the flow's special structure and how it depends on friction and other factors. In this section, we describe how our experiments allow us to build on these basic ingredients to identify two qualitatively different regimes in vibration dominated flows. We find a frictional regime in which $P \sim \tau$, yet with $\mu(\Omega)$ a rate dependent friction for $\Omega \gtrsim$ 10$^{-3}$~rps or $\Gamma \lesssim 0.8$. For even slower, more strongly vibrated flows, both the rheology and the location of the shear band presents strong evidence for a regime where $T$ becomes \emph{independent} of $P$.

\begin{figure}[t!]
    \begin{center}
				\includegraphics[width=\columnwidth]{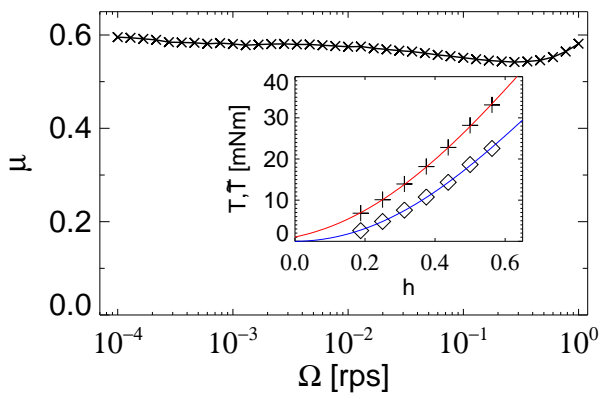}
        \caption{(Color online) $\mu(\Omega)$ as obtained from the fit with the frictional model. The inset shows one example of the fit for $\Omega=8.5\cdot\times10^{-4}$~rps. The upper curve (+) shows the raw data $T$, the bottom curve ($\diamond$) is the raw data minus the correction term, $\tilde{T}$ -- which goes through the origin. The fit matches the data very well resulting in a $\chi^2$ of 2.0$\times 10^{-3}$ (for the upper curve).}\label{fig:g0metinset}
    \end{center}
\end{figure}

\subsection{Torque Minimization Model}\label{sec:unger}
To interpret the observed filling height dependence of the shear stresses, we start from a simple frictional model due to Unger {\em et al.}, which was developed to describe the 3D shape of the shear zones in the split-bottom geometry, but which also makes a precise prediction for the driving torque as function of filling height for purely frictional flows \cite{2004_prl_unger,2006_prl_fenistein,2010_sm_dijksman}. This model is based on the following three ingredients. First, think of the shear zones as localized along a narrow sheet $r(z)$ (corresponding to the center of the shear zones \cite{2003_nature_fenistein,2004_prl_fenistein, 2006_prl_fenistein, 2010_sm_dijksman}). Second, assume that the stress tensor is colinear with the strain rate tensor \cite{2006_pre_depken} and proportional to the hydrostatic pressure. Third, assume that the sheet shape $r(z)$ minimizes the driving torque \footnote{Note that whereas we characterize the filling height with the dimensionless $h$, we write the frictional model in terms of $H$, the common notation.}:
\begin{equation}
\tilde{T}[r(z)] = 2 \pi g \rho \mu \int^{H}_{0}(H-z)~r^2\sqrt{1+(dr/dz)^2}~dz ~, \label{ttildeH}
\end{equation}
where $g$ denotes the gravitational acceleration, and $\rho$ the bulk density (1.7$\times$10$^3$ kg/m$^{3}$) of the granular material. Minimizing $\tilde{T}$ for a given $h$ determines the shear sheet $r(z)$, from which the torque can be determined as function of $h$. As expected, we can write this torque as $\tilde{T}(h)=\mu \tilde{T}_f(h)$, where $\tilde{T}_f$ is a universal function of $h$. Note that for shallow filling heights, the torque is approximately proportional to the product of pressure and the extension of the shear band, so that $\tilde{T}_f(h)$ is quadratic in $h$ for $h \ll 1$.

In contrast to the original split-bottom cell for which Eq.~(\ref{ttildeH}) was developed, in our system the driving disk is slightly elevated with respect to the bottom. This is done in order to avoid observing spurious torque fluctuations that we associate with the diverging strain rate in the original split-bottom setup. The elevated disk leads to a $\mu$ depended addition in the experimental torque signal $T$, due to slip between the side of the disk and the stationary particles next to it. We have found that this drag term can be estimated as:
 \begin{equation}
    T_{drag}(H)=2\pi r_s^2\mu \rho g \int_0^s (H+z)dz~, \label{drageq}
\end{equation}
where $s$ is the disk thickness (5~mm), and $\mu$ is the effective friction coefficient for sliding of the disk past the particles, for which we use the same effective friction coefficient as for the granular flow. The contribution of $T_{drag}$ to the torque varies with $H$ and is proportional to $\mu$, so that we can write $T_{drag} = \mu T_d$, where $T_d$ can be deduced from Eq.~(\ref{drageq}).

We conclude that the measured torque $T$ is composed of two contributions:
\begin{equation}
T=\tilde{T}(H)+T_{drag}= \mu \left[ \tilde{T}_f(H) + T_d\right]=\mu T_f~. \label{ungereq}
\end{equation}

In conclusion, we can extract $T(h)$ from our flow curves, and check whether the flow appears frictional, and if so, determine $\mu$ \cite{2010_pre_dijksman} and $\tilde{T}(h)$.

\subsubsection{$\Gamma$=0}
In Fig.~\ref{fig:overview_h}a we show flow curves for $\Gamma=0$ and a range of $h$. Clearly, the torque only weakly varies with $\Omega$, and we expect the stresses to be frictional. For each fixed $\Omega$, we extract $T(h)$ from our data and fit it to $\mu T_f$ (Eq.~(\ref{ungereq})), as shown in the inset of Fig.~\ref{fig:g0metinset}. We find that this fit is excellent, which implies that the stresses are frictional, and which allows us to extract  $\mu(\Omega)$. As shown in Fig.~\ref{fig:g0metinset}, $\mu(\Omega)$ is almost flat, and has the same shape as the flow curves. We stress here that $\mu(\Omega)$ together with the frictional model predicts the stresses for \emph{all} values of $h$, thus representing all the flow curves taken at different $h$.  We note that our values  for $\mu$ are comparable to those found previously in a standard split-bottom cell using the same particles~\cite{2010_sm_dijksman}.

\subsection{Frictional Model for $\Gamma>0$}\label{sec:Ggt0}
In Fig.~\ref{fig:overview_h}b and Fig.~\ref{fig:overview_h}c we show examples of flow curves for a range of $h$ and $\Gamma>0$. We will now use $T(h,\Gamma>0)$ to test if the basic assumptions for the Unger model break down in the vibration dominated regime. We will find two flow regimes with the distinguishing features $T \sim \mu(\Omega) P$ and $T \nsim \mu P$. We describe here how we can distinguish these regimes in the rheological data.

\begin{figure}[t!]
    \begin{center}
        \includegraphics[width=\columnwidth]{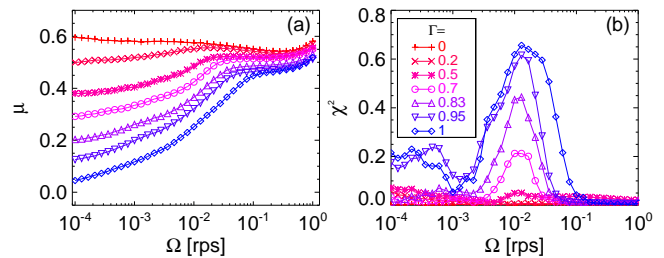}
        \caption{(Color online) (a) The effective friction coefficient $\mu(\Gamma,\Omega)$ as found by fitting the data with Eq.~\ref{ungereq}. (b) The $\chi^2$ of the fits, as defined in Eq.~(\ref{chi2}). }\label{fig:3x3}
    \end{center}
\end{figure}

From Fig.~\ref{fig:overview_h}b-c we see that that the flow curves for $\Gamma>0$ all show significant rate dependence. Even so, we attempt to fit Unger's model to the rheological data. We thus fit $T(h,\Omega)$ to try to obtain a $\mu(\Omega)$. If this rate dependence were captured by an effective friction coefficient that only depends on $\Omega$, with $T(h,\Omega) = \mu(\Omega) T_f(h)$, the rate dependence would lead to a good fit of our data to the frictional model. To quantify the deviations between the data and fits to the frictional model, we calculate the best estimate of $\mu$ and the corresponding $\chi^2_f$ as follows. For each fixed $\Omega$ and $\Gamma$, we have measured the torque for seven values of $h$, and then determine:
\begin{equation}\label{chi2}
\chi^2 := \langle (\mu T_f(h) - T(h))^2 \rangle / \sigma^2_{T(h)}~,
\end{equation}

We apply this procedure for each value of $\Omega$ and $\Gamma$, and show the result for $\mu$ and  $\chi^2$ of these calculations in Fig.~\ref{fig:3x3}. For $\Omega>0.3$~rps - the inertial regime - the fit works very well and results in a weakly rate dependent $\mu$, just as for $\Gamma=0$. In addition, we find a large region for $\Gamma \leq 0.83$ and $\Omega<0.5\times 10^{-3}$ where the fit also works well, but this time with a more strongly rate dependent effective fiction $\mu(\Omega)$. This tells us that even in this rate dependent, vibration dominated regime, a frictional prediction is perfectly capable of describing the flow.

We do however observe two distinct regimes where $\chi^2$ is large, indicating a poor fit. First, there is a significant peak in $\chi^2$ around $\Omega=10^{-2}$~rps for $\Gamma\geq0.7$. Second, for $\Gamma\geq0.95$ and $\Omega<10^{-3}$~rps, $\chi^2$ also is substantial. The underlying physics in these two regimes is different. As we will show in the next section, the first peak is associated with a broad crossover regime between rate dependent and rate independent flows -- a rather trivial consequence of the flow profiles in the split-bottom geometry. The second peak we associate with a flow regime in which the rheology becomes pressure independent, as shown in in section~\ref{sec:fluid}.

\begin{figure}[t!]
    \begin{center}
        \includegraphics[width=\columnwidth]{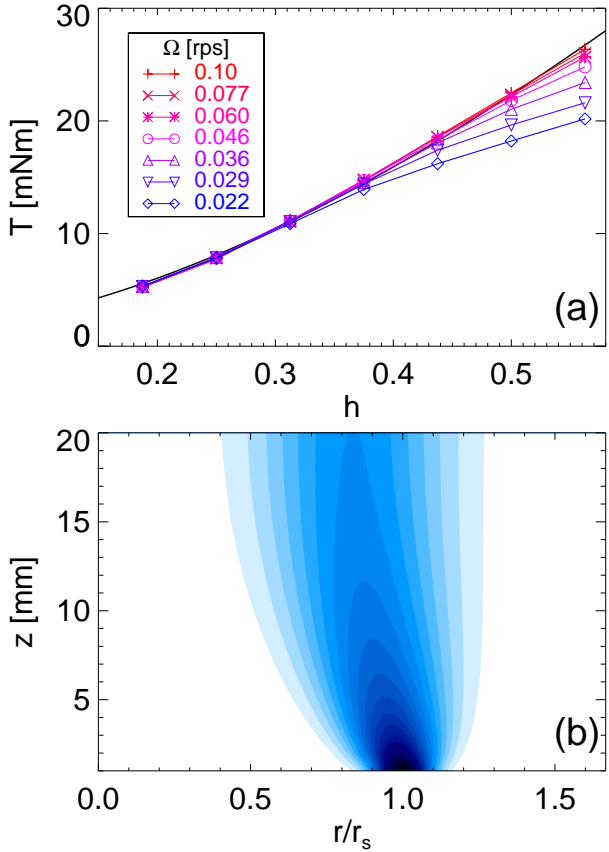}
        \caption{(Color online) (a) $T(h)$ for $\Gamma$=0.95 and a range in $\Omega$ at the point where the rate dependence starts. It can be seen that the curve drop for high $h$, resulting in an s-shaped $T(h)$ curve rather than an upwards curved $T_f$ one. The black line is the fit with the frictional model to the top curve. (b) A theoretical prediction of $\dot{\gamma}/\Omega$ -which decreases with $z$- in the split-bottom cell~\cite{2010_sm_dijksman,wandesman_epl_2012}. Dark color presents high $\dot{\gamma}/\Omega$, white is low $\dot{\gamma}/\Omega$.}\label{fig:ratedep}
    \end{center}
\end{figure}

\subsubsection{Onset of Rate Dependence}
The peak in $\chi^2$ around $\Omega=10^{-2}$~rps is consistent with the onset of rate dependence below $\Omega =10^{-2}$ rps as per the following reasoning. First, both our raw data for $T$ as well as the best fits for $\mu$ show that rate dependence sets in rather abruptly for $\Omega<0.1$~rps, and that rate dependence is strongest for large $\Gamma$, consistent with the location and strength of the peak in $\chi^2$. Crucially, this onset of rate dependence sets in at different flow rates for different heights (see Fig.~\ref{fig:overview_h}), so that at a given $\Omega$, the data for $T(h)$ mixes rate independent and rate dependent flows.

In Fig.~\ref{fig:ratedep}a we show examples of $T$ as function of height, that illustrate that when $\Omega$ enters this rate dependent regime, $T(h)$ strongly deviates from the quadratic form predicted by Eq.~(\ref{ungereq}). To interpret this deviation, it is important to realize that at a given $\Omega$, the local strain rate $\gammadot$ spans a wide range of values and has a strong $z$ dependence
\cite{2004_prl_unger,2010_sm_dijksman,wandesman_epl_2012} --- see Fig.~\ref{fig:ratedep}b. Hence, as the torque $T$ is an integral over the local stress in different layers in the material, $T(\Omega)$ mixes different local rheologies. More precisely: under the assumption that
rate dependence sets in below a given $\gammadot$, there is a range of values of $\Omega$ for which the lower part of the system (where strain rates are largest) is
still rate independent, whereas the top part of the system (where strain rates are smallest) are already rate dependent. This is consistent with the ''drop'' in the $T(h)$ curves at large $h$ shown in Fig.~\ref{fig:ratedep}a --- the deviations from the Unger model emerge first for large $h$, for which the range of strain rates is biggest and regime mixing is thus most pronounced.
Our data also shows that once $\Omega$ is sufficiently low, so that all of the material is in a rate dependent state, $T(h,\Omega)$ is close to  $\mu T_f(h)$ so that $\chi^2$ drops to low values again, and $\mu$ can be replaced with a rate dependent $\mu(\Omega)$. The range of $\Omega$ over which this crossover exists broadens with $\Gamma$, since the rate dependence becomes stronger with $\Gamma$.

In conclusion, the lowering of the friction coefficient $\mu$ and the peak in $\chi^2$ around $\Omega=0.01$~rps are caused by the onset of rate dependence which occurs at different $\Omega$ for different vertical locations in the flow cell. For $\Gamma \lesssim 0.8$, we also observe that once all the material is in the slow, rate dependent regime, the fit to the frictional model achieves a low $\chi^2$ again, so that $T(h,\Omega) \approx \mu(\Omega) T_f(h)$.

\begin{figure}[t!]
    \begin{center}
				\includegraphics[width=\columnwidth]{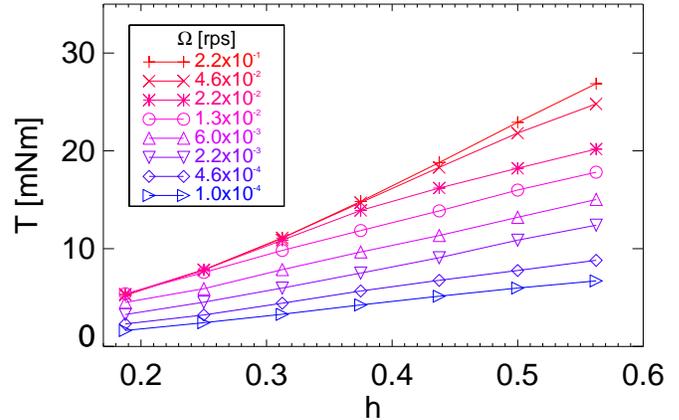}
        \caption{(Color online) $T(h)$ curves for $\Gamma=1$ and $\Omega<10^{-3}$~rps. Towards lower $\Omega$, the curves lose their curvature and become straight lines. }\label{fig:toline}
    \end{center}
\end{figure}

\subsection{Fluidized Region}  \label{sec:fluid}
The growth of $\chi^2$ for large $\Gamma$ and low $\Omega$ signals a breakdown of the frictional picture, where shear stresses are proportional to the pressure, as we will describe in this subsection.  To gain deeper insight in the flow phenomenology in this regime, we plot $T(h)$ for $\Gamma=1$ and a range in $\Omega$ in Fig.~\ref{fig:toline}. We see that for all filling heights the stresses drop with $\Omega$, and at low $\Omega$, $T(h)$ becomes approximately {\em linear}. The standard Unger model predicts a quadratic dependence of $T(h)$ on $h$, as mentioned above. A linear dependence would suggest a pressure independent rheology, for which the increase of $T$ with $h$ is only due to increasing surface area on which the shear stress acts.

Note that the large values of $\chi^2$ here cannot be due to the existence of a crossover regime, as presented above for $\Omega \approx 10^{-2}$~rps. Evidence for this comes from Fig.~\ref{fig:overview_h}c, which shows that both rate dependence of $T(\Omega)$ is small, and that there is no strong difference in the rate dependence for different values of the height in this regime. First of all, that means that there is little mixing of different rheologies in the global torque signal; second, the rate dependence is weak, so even if there were some mixing, it would not produce a strong $h$ dependence.

It it perhaps not surprising that new phenomena occur around the special value $\Gamma=1$. For $\Gamma \approx 1$, the grains lose contact during part of the vibration cycle --- the precise value of $\Gamma$ where this happens depends on details \cite{durian_prl_1997,2005_pre_mobius}. As a result, the confining pressure becomes zero during part of the cycle, and as most slip can be expected to  occur when the normal grain forces are absent, the flows may become pressure independent, as in a viscous liquid.

\begin{figure}[t!]
    \begin{center}
				\includegraphics[width=\columnwidth]{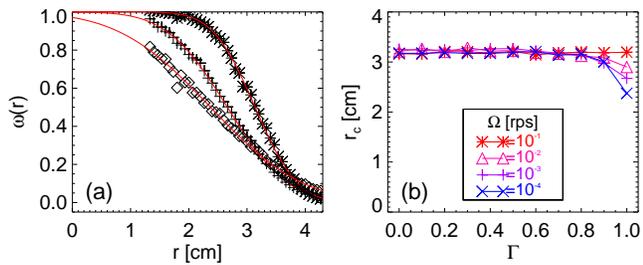}
        \caption{(Color online) (a) $\omega(r)$ for $\Omega=10^{-1}$~rps, $\Gamma=0.2$ ($\times$), $\Omega=10^{-3}$~rps, $\Gamma=1$ (+),  $\Omega=10^{-4}$~rps, $\Gamma=1$ ($\diamond$). In red, we add the fit with $\omega(r)=1/2-1/2$~erf$[(r-r_c)/W]$. (b) The center of the shear band at the surface $r_c$ as a function of $\Gamma$ for h=0.47 and $\Omega$ as indicated. For low $\Omega$, $r_c$ clearly decreases.}\label{fig:flowprofile}
    \end{center}
\end{figure}
\subsubsection{Rate Dependent Flow Structure}
Additional evidence for the loss of pressure dependence for high $\Gamma$ and low $\Omega$ comes from measurements of the flow structure. From finite element calculations on the flow structure of a viscous liquid in the split-bottom geometry, it is known that the shear band is much closer to the center of the cell than for frictional flow~\cite{2010_pre_dijksman}. As such, a pressure independent rheology for the granular flows in this regime can be expected to be accompanied by similar changes in the flow structure.

To test this, we have measured the velocity profiles $\omega(r)$ at the surface of our system for a range in $\Gamma$ and $\Omega$ using particle image velocimetry \cite{2004_prl_fenistein, 2006_prl_fenistein, 2010_sm_dijksman}. In Fig.~\ref{fig:flowprofile}a we show examples of $\omega(r)$, showing a broadening and shift of the shear zones when $\Omega$ enters the pressure independent regime.
We fit the velocity profiles with $\omega(r)=1/2-1/2$~erf$\left[ (r-r_c)/W \right]$, where $r_c$ is the center of the shear band at the free surface \cite{2004_prl_fenistein}. In Fig.~\ref{fig:flowprofile}b, we plot $r_c$ for $\Omega$ ranging from $10^{-1}$ to $10^{-4}$ ~rps and a range in $\Gamma$. Clearly, the location of the shear band is mostly independent of $\Omega$ and $\Gamma$, including in most of the rate dependent regime. However, in the regime where we observed the pressure independent rheology, we observe significant deviations in the flow profiles. The deviations show a trend towards a shear band moving inwards --- consistent with the idea of a pressure independent regime.

\begin{figure}[t!]
    \begin{center}
				\includegraphics[width=\columnwidth]{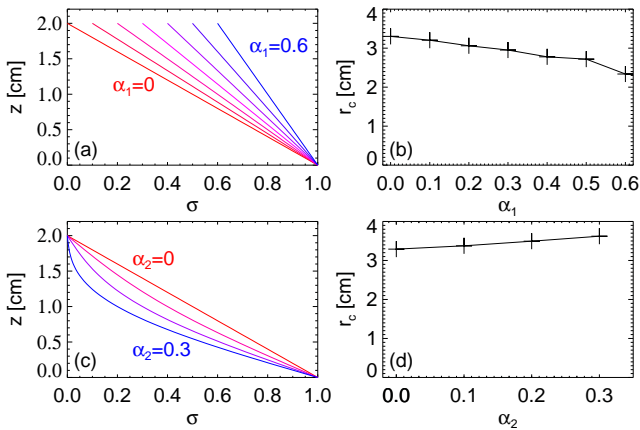}
        \caption{The center of the shear band at the surface, $r_c$, can be found using the method by Unger for given $z$ dependent stress $\sigma(z)$. In (a), we plot $\sigma(z)$ for case $\sigma(z) \sim 1-(1-\alpha_1)z/H$ where $\sigma$ is still finite at $z=H$, in contrast to the frictional description where $\sigma(H)=0$. The resulting $r_c$ is shown in (b), we recover in inwards moving shear band as we observe in experiments. In (c) we plot $\sigma(z)$ for $\sigma=(1-z/H)+\alpha_2\sin(2 \pi z/(2H))$, where $P$ and $\mu$ respectively vanish and decrease towards the surface, corresponding the a strain rate dependent frictional picture, as we show in (d), this predict an outwards moving shear band, contrary to what we observe.}\label{fig:ungershift}
    \end{center}
\end{figure}

Moreover, we can modify the Unger model to test which rheological scenario is most compatible with the observed shift in the shear band. Throughout, we assume that the torque minimization principle is robust. The frictional torque model assumes $\sigma(z) \sim 1-z/H$, in which the shear stress, being proportional to the hydrostatic pressure, goes to zero at the surface. We can replace this model with $\sigma(z) \sim 1-(1-\alpha_1)z/H$, in which the shear stress reaches a final value when approaching the free surface -- see Fig.~\ref{fig:ungershift}a. The extreme case $\alpha_1 = 1$ represents a pressure independent rheology. We compute the location of the shear band at the free surface as a function of model parameter $\alpha_1$. The results are shown in Fig.~\ref{fig:ungershift}b. We find that for larger $\alpha_1$, the location of the shear band at the free surface moves inwards. Thus, the closer the model resembles a Newtonian rheology, the more the shear band moves towards the center. This can be understood intuitively as follows: the penalty for having a shear band at large radius at the surface is zero in the pressure dependent model, because the shear stress goes to zero at the free surface. Once a finite amount of shear stress is present in the shear band at the surface, torque minimization will move the shear band inwards precisely as we observe in the experiments at $\Omega < 10^{-3}$~rps, $\Gamma > 0.9$.

Conversely, for a frictional, rate dependent rheology, the shear stress closer to the surface is \emph{lower} than that of a simple frictional model. We model this with a $\sigma(z)$ that can be captured with $\sigma=(1-z/H)+\alpha_2\sin(2 \pi z/(2H))$, as shown in Fig.~\ref{fig:ungershift}c. The torque penalty for having a shear band at finite $r$ is thus reduced, and the model predicts indeed an \emph{increase} of the shear band radius at the free surface (Fig.~\ref{fig:ungershift}d), contrary to what we observe. We thus conclude that our observation of the inward displacement of the shear band location at $\Omega < 10^{-3}$~rps, $\Gamma > 0.9$ is consistent with the granular flow obtaining a rheology which becomes pressure independent.

\section{Constant Torque Driving} \label{sec:constantT}
\begin{figure}[th!]
    \begin{center}
        \includegraphics[width=\columnwidth]{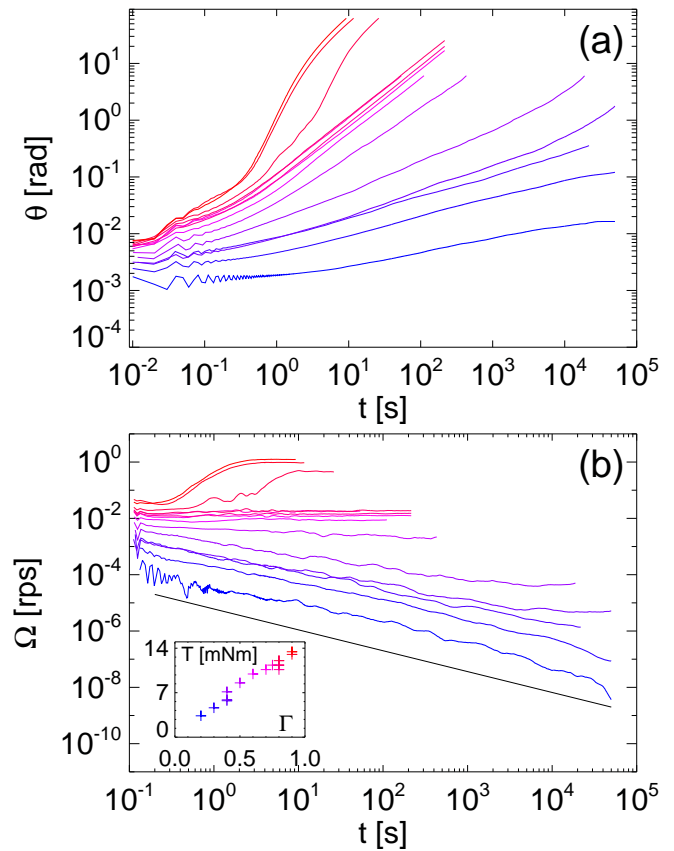}
        \caption{(Color online) The phenomenology for constant torque experiments at $H/R_s = 0.33$.
        In (a), we show the $\theta$ as a function of time, in (b), we show the derivative of $\theta$: $\Omega$. It can be seen that the curves for low $\Gamma$ and $\tau$ show very long transients. The black line has a slope -1. The curves are down up for $\Gamma=$0.2,0.3,0.4,0.4,0.4,0.5,0.6,0.7,0.75,0.8,0.8,0.8,0.9,0.9,1 and $T$~[mNm]=3.36, 4.62, 5.74, 5.88, 7.14, 8.54, 9.94, 10.64, 11.3, 10.64, 11.3, 12.04,13.02,13.44, as also indicated in the inset of (b).}\label{fig:Tmerge}
    \end{center}
\end{figure}

In this section we discuss stress controlled experiments. These experiments consist of the following protocol: after pre-shearing the sample, a fixed value of
the driving torque $T$ is set, and the ensuing rotation angle $\theta$ of the disk is monitored. We show several examples of the time evolution of $\theta$ and its derivative $\Omega$ for a range of values of the torque $T$ and vibration strength $\Gamma$ and fixed $h = 0.33$ in Fig.~\ref{fig:Tmerge}. Data for other values of $T$, $\Gamma$ and $h$ look similar; the small initial oscillatory motion, visible for small $\Gamma$, is due to transient flexure oscillations.

As shown in Fig.~\ref{fig:Tmerge}, in most cases the disk rotation rate reaches a steady state value after some equilibration time. For small $T$, this equilibration time becomes extremely long, much longer than in the rate controlled experiments shown above. This is typical for complex fluids that exhibit a so-called viscosity bifurcation \cite{coussot_pre_2002,bonn_epl_2009}, and we suggest that weakly vibrated granular media fall into this category. We also note that for small $T$, the angle as function of time (or alternatively, $\Omega(t)$) exhibits power law behavior reminiscent of Andrade creep \cite{andrade_1914}, again a typical feature of weakly driven disordered media.

For the data sets where $\Omega$ reaches a steady state, we find that $\Omega(T)$ and $T(\Omega)$ are consistent --- in other words, the steady state flow properties are the same as measured through the rheology under constant rate or constant stress driving. Consistent with this picture, we observe that the data in Fig.~\ref{fig:Tmerge} illustrates that the steady state values of $\Omega$ exhibit a jump between inertial rates ($\Omega>0.5$~rps) and significantly slower flow rate ($\Omega < 10^{-2}$~rps), consistent with the hysteretic transition between inertial and vibration dominated flows observed in \cite{2011_prl_dijksman}.

\section{Conclusion}\label{sec:conc}
To summarize, we probe the rheology of weakly vibrated granular media and find several different flow regimes. First, for $\Omega>0.3$~rps, our data shows the well known \emph{inertial flow} regime, consistent with a rough estimate of the inertial number. In the absence of vibrations, lower flow rates lead to an essentially rate independent, \emph{quasistatic}, regime, where the variation of the torque is small, and where  $T(h)$ is well fitted using Eq.~(\ref{ungereq}), implying that the shear stresses are proportional to the pressure here. For $\Gamma=0$, this regime covers all $\Omega < 0.1$~rps, whereas the range of flow rates where this  rate independent regime resides  shrinks in the presence of vibrations, and almost vanishes for $\Gamma=1$. For $\Omega$ below the rate independent regime and $\Gamma>0$, we have described three \emph{vibration dominated} regimes. For two of these regimes, our data shows that the shear stresses are still proportional to the normal stresses, but now via rate dependent $\mu(\Omega)$.  For the slowest of these two regimes, we see a slow densification, leading to a kink in the flow curves.  Finally, for $\Gamma$ close to one, the vibrations affect the rheology of the granular medium so significantly, that the  shear stresses are no longer proportional to the normal stresses, signifying a complete departure of the frictional nature that is a hallmark of all other types of slow granular flows.

\section*{Acknowledgements}
We appreciate helpful discussions with O. Dauchot and H. Jaeger, J. Mesman for outstanding support in constructing the setup, and T. Tampung for support in developing the electronics. We thank L. van Dellen for his contribution to the experiments and data analysis. This work was supported by the Dutch physics foundation FOM.

\newpage
\appendix
\section{Setup} \label{app:setup}
\begin{figure*}[t!]
    \begin{center}
        \includegraphics{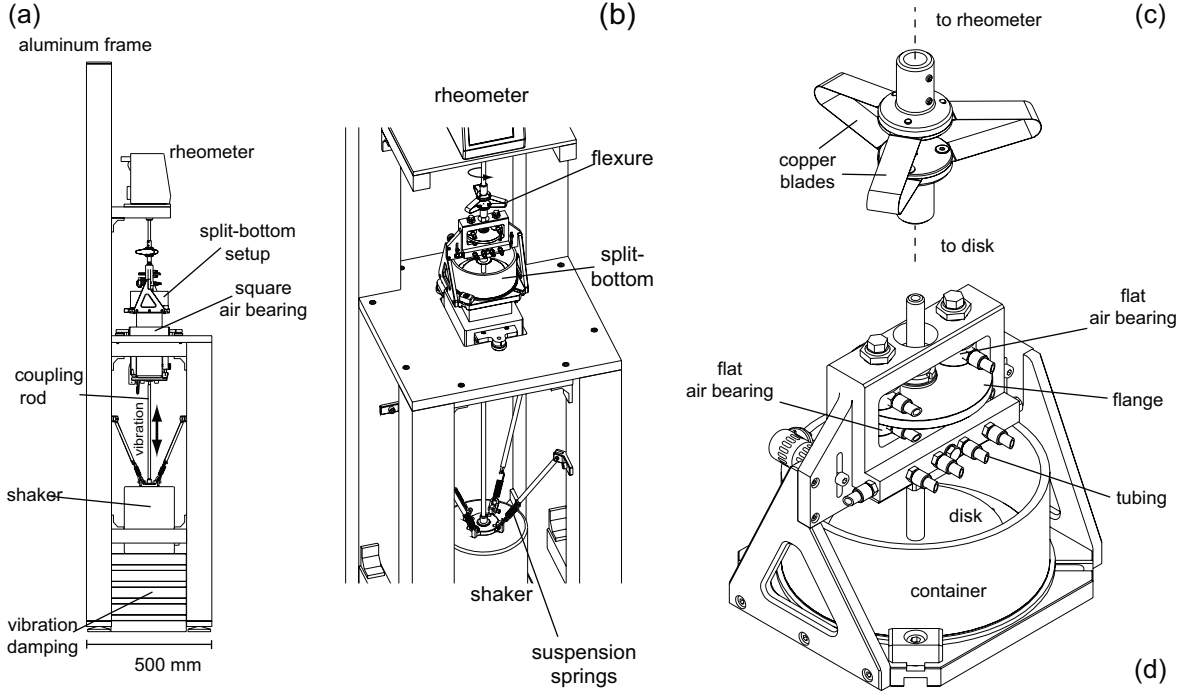}
        \caption{\label{fig:setup}
		The setup used in the vibrated-rheology experiments. (a) Side-view of the setup. (b) Perspective view of the split-bottom cell on the air bearing, with the shaker beneath it. (c) The flexure coupling the rheometer axis to the disk axle. The copper blades measure 0.3~mm in thickness and 7~mm in width. (d) The split-bottom cell, with the flat air-bearing assembly. The four flat bearings that clamp the flange and fix it rigidly to the container are mounted on a bridge over the container itself. The round bearing that holds the axis for the disk is also mounted in the bridge. Tubing for compressed air is not shown.}
    \end{center}
\end{figure*}
Our setup allows to do sensitive rheology on vibrated materials and its main parts are illustrated in Fig.~\ref{fig:setup}. The granular split-bottom flow cell forms the heart of setup, and consists of a cylindrical flow cell, the bottom of which is formed by a rotating disk that drives the flow. The disk is attached to an axle which is mounted in a cylindrical air bearing ($\varnothing$ 1/4",
New Way): rotational motion is friction-free which allows precise rheological measurements.
\begin{figure}[h!]
    \begin{center}
        \includegraphics[width = \columnwidth]{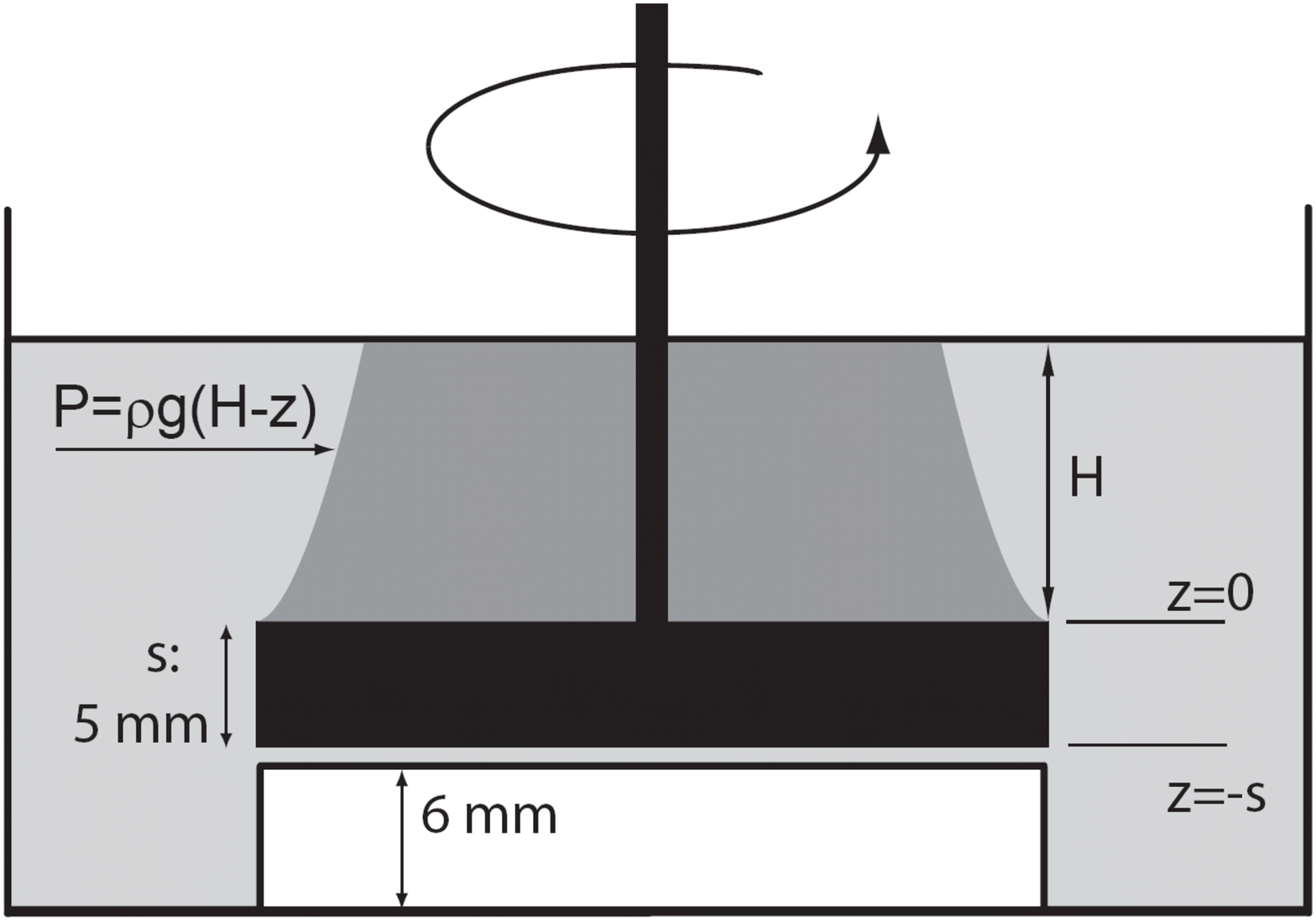}
        \caption{Schematic side view of the split-bottom setup. A cylinder of height 6~mm is placed underneath the disk, which has a thickness $s$ of 5~mm and a radius $r_s$ of 40~mm. The gray area represents the volume occupied by the grains; the dark grey region indicates the volume of particles co-rotating with the disk in the trumpet regime. The hydrostatic pressure $P$ acts on the interface between the co-rotating and `static' volume of particles as indicated. The side of the disk is also exposed to particles and, although smooth, contributes to dissipation during rotation as discussed in the text.}\label{fig:diskfric}
    \end{center}
\end{figure}

In ordinary split-bottom cells, the strain rate diverges at the split. We have found that
the torques occasionally show strong fluctuations in this case, presumable due to individual particle being trapped just above the split --- moreover, these fluctuations depend on the precise roughness near the split, thus leading to a dependence of the average $T$ on such experimental details.
To avoid this, we make sure that the strain rate field is smooth at the grain level and have elevated the driving disk by 6~mm  (Fig.~\ref{fig:diskfric}). The side of the disk is smooth, and particles immediately next to the disk hardly move, creating a static bottom layer flush with the disk. Hence, the boundary conditions are essentially the same as for the ordinary split-bottom disk, and the elevation does not affect the overall flow field for $h\equiv H/r_s$ larger than about 0.1. The elevation does ensue that the torques are insensitive to experimental details and do not show the aforementioned spurious fluctuations.

We induce vibrations in the split-bottom cell with an electromagnetic shaker (VTS systems VG100), driven by a function generator (Thurlby Thandar TG1010) via a commercial audio amplifier (Crown CE1000). The shaker is coupled to the flow cell via a long rod, and the cell is mounted on a square air bearing (4"x4", New Way), which ensures that the slider and flow cell move virtually friction free in the vertical direction, with the amplitude of the residual horizontal vibrations in the submicron range due to the stiffness of the air bearing. The total combined weight of the slider and the split-bottom cell is approximately 12~kg and too large to be carried by the shaker alone. Therefore we support their weight with four suspension springs with a stiffness of 2~kN/m, connected to the aluminum frame shown in Fig.~\ref{fig:setup}a,b. The vertical acceleration of the split-bottom cell is measured by a combination of two accelerometers (Dytran E3120AK and a modified ADXL322EB~\footnote{Its factory default bandwidth of 50~Hz is increased to 2~kHz.}). To limit the mechanical coupling between the table and the shaker, the shaker is placed on a stack of concrete tiles with rubber mats sandwiched between them. We do not observe any appreciable heaping~\cite{1995_prl_pak}.

\begin{figure}[t!]
    \begin{center}
        \includegraphics[width=\columnwidth]{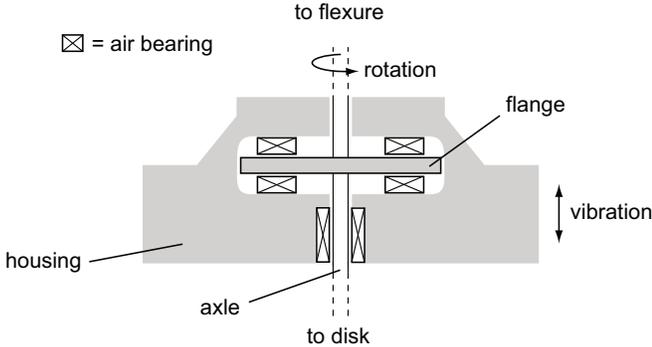}
        \caption{Schematic drawing of the custom-built air-bearing assembly that
        fixes the disk axle in the vibrating reference frame of
        the container. The flange is held in place by four flat bearings,
        the axle is fixed by a cylindrical bearing.\label{fig:cusairbearing}}
    \end{center}
\end{figure}

The vibrations in the flow cell should not induce relative vertical motion between the disk that drives the granular flow and the container. To ensure that the disk only induces rotational shear deformations and no additional vibrations, we attach a flange to the disks axle. The surface of the flange is polished, and clamped between four flat air bearings ($\varnothing$ 25~mm, New Way) -- see Fig.~\ref{fig:cusairbearing}. By mounting the flat air-bearing assembly on the container, we ensure that disk and container move together during the vibration cycle: the stiffness of one flat bearing, for a typical gap between the bearing and the flange of 5~$\mu$m, is 18~N/$\mu$m. Since the peak acceleration of the system is never larger than $\sim 1.5$~$g$, and the weight of the flange/axle/disk construction is less than a kilogram, the maximum residual motion between the disk and the container is smaller than a micrometer, which is the typical length scale of asperities on glass beads~\cite{2005_pre_schroter}.

A rheometer (Anton Paar DSR 301), coupled to the flow cell via a system of flexures and airbearings, allows to probe and control the granular flow. As the rheometer cannot be vibrated, and its rotational axis is housed in a very stiff air-bearing, we need to decouple the vibrations from the rheometer, and also compensate for inevitable misalignments of rheometer and flow cell. We therefore couple the two axes with a flexure, as shown in Fig.~\ref{fig:setup}c. This flexure has a torsional spring constant of 4~Nm/rad and compressional spring constant of 5~$\times 10^2$~N/m. The flexure can be compressed easily and therefore accommodate the vertical motion of the disk axle. The copper blades used in the flexure are flexible enough to compensate for any small misalignment between the axis of the disk and the rheometer, without exerting significant forces on the bearings.

Our DSR 301 rheometer is a stress driven rheometer, so its native mode is to set a torque to control rotation. However, it is also possible to use the rheometer in a so-called constant shear rate mode. In this mode, the rheometer applies a feedback routine that dynamically adjusts its torque in order to achieve a constant rotation rate. This process has one control parameter: the rate at which the torque is increased or decreased. This timescale is set by the desired driving rate, and by a user-defined time constant, and in principle could affect all constant shear rate experiments. In these experiments we set the feedback constant such that we observe smooth rotation, which is achieved for a feedback parameter, the so called CSR-value, of 25\%. If the torque adaptation rate is too high and becomes of the order of the eigenfrequency of the rheometer axle, the feedback routine will be affected by the rotational eigenmodes of the mass-spring system of the rheometer axle.

\subsection{Alignment}

To compensate for inevitable misalignments between the shaker axis and vertical air-bearing axis, we couple them via a 40~cm long and flexible aluminum rod. The aluminum rod is flexible enough to compensate for a horizontal alignment mismatch of a few millimeter, but stiff enough to ensure a mechanical coupling for vertical motion (see Fig.~\ref{fig:setup}a).

Misalignment between the disk and rheometer axle causes rotationally periodic variations in the force that
the rheometer exerts on the disk and grains. This periodic modulation of the torque can be reduced by improving the alignment of the flexure-coupled axles. We manually align the rheometer with the disk axis by setting a constant rotation rate to the disk, with no particles in the container. While the disk rotates, we monitor the required torque to run the disk at the set speed. This torque has a periodic modulation; alignment of the two axis with the set screws on the air bearing reduces the amplitude of this periodic modulation on the torque.

\begin{figure}[t!]
    \begin{center}
        \includegraphics[width=6cm]{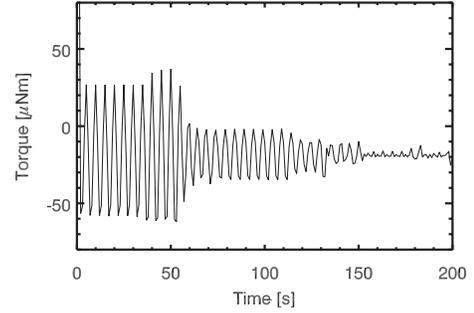}
        \caption{A typical torque signal under constant
        rotation rate during the alignment of the air bearing. One oscillation period corresponds
        to a full rotation of the disk}\label{fig:residue}
    \end{center}
\end{figure}

A typical alignment run is shown in Fig.~\ref{fig:residue}: the modulations can be minimized to an amplitude of less than 20~$\mu$Nm, which is $<$ 0.1\% of the largest torque need to drive our granular flows. Nota that a small torque offset also remains since the flat air bearings that confine the vertical motion of the flange cannot be positioned perfectly parallel to the flange. This induces an asymmetry in the air flow between the bearings and the flange. This exerts a torque of the order of 20~$\mu$Nm on the flange; a negligible amount in comparison to the typical driving torques employed in our experiments.

\subsection{Vibration Control}

\begin{figure}[t!]
    \begin{center}
        \includegraphics[width=6cm]{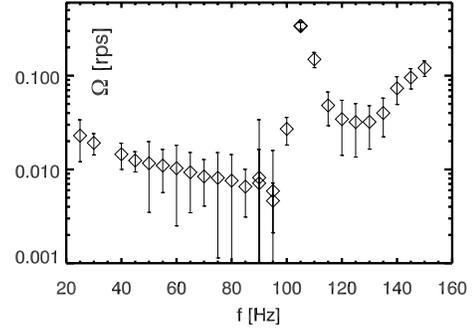}
        \caption{Steady state rotation rate  dependence on the vibration
        frequency, measured with $\Gamma = 0.5$, $\tau = 0.8$.
        }\label{fig:vibrheo:mechresp}
    \end{center}
\end{figure}

To choose an appropriate external vibration frequency, we measure how the steady state rotation
rate $\Omega$ under a constant applied torque $T=25$~mNm and fixed vibration amplitude $\Gamma = 0.5$ depends on the vibration frequency $f$. We carry out the measurements with a filling height of $h$ = 0.55. The result is shown in Fig.~\ref{fig:vibrheo:mechresp}. We see that there is a resonance above 105~Hz, and for vibration frequencies between 50 and 75~Hz, the rotation rate depends only marginally on the vibration frequency. We choose the middle of this regime, $f$ = 63~Hz to carry out all experiments discussed in this paper. The vertical vibration amplitude at a peak acceleration of 1~$g$ at this frequency is 64~$\mu$m.

It is critical that the vibration amplitude is constant during the duration of the experiments, which can last for hours. The shaker generates considerable heat, which affects its operation; with constant driving current to the shaker coils, the vibration amplitude slowly drifts over $\sim 0.1 g$ per hour.

To keep the vibration amplitude constant during the experiments, we employ a LabVIEW driven feedback scheme that continuously adjusts the driving current for the shaker coil. A function generator, under computer control, sends a sine wave to the amplifier that drives our shaker. Accelerometers detect the actual shaking amplitude. We find that the harmonic distortion of the sinusoidal shaker motion is about 1\%, so we use a lock-in amplifier (SRS 830) to achieve a high accuracy measurement of the vibration amplitude. The computer then adjust the output of the function generator, and with this feedback scheme we can maintain a stability of $\Gamma$ to within $<10^{-3}$~$g$.

\subsection{Steady state $T(t)$} \label{proto}
\begin{figure}[t!]
    \begin{center}
        \includegraphics[width=\columnwidth]{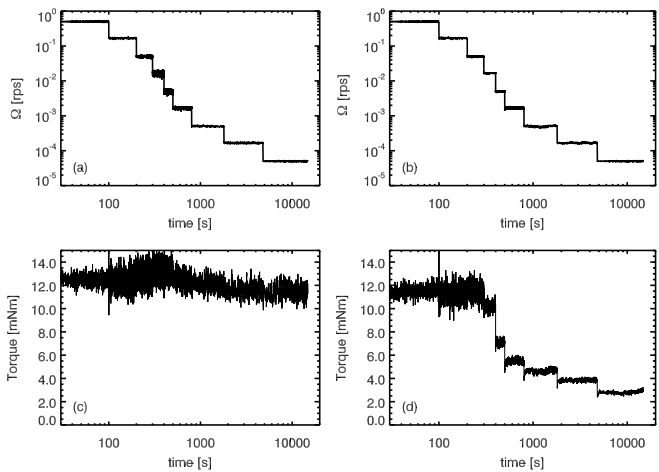}
        \caption{Rotation rate (a,b) and driving torque (c,d) shown as a function of time for an experiment in which the set speed is reduced logarithmically. (a,c) are measured for $\Gamma = 0.2$, (b,d) are measured for $\Gamma = 0.8$. Both experiments are carried out at $h = 0.33$.}\label{fig:transient}
    \end{center}
\end{figure}

We verify that a steady state flow rate can be obtained by the rheometer and that we observe no long term transients in the rate control mode. To do this, we measure the instantaneous driving torque and speed as a function of time for different rotation rates $\Omega$ and vibration amplitudes $\Gamma$. We apply pre-shear before the start of the experiment, and start a logarithmic ramp down in $\Omega$, from $5\times 10^{-1}$ to $5\times 10^{-5}$~rps with 2 steps per decade. We measure for all $\Omega$ over a minimum strain of 0.5 rotation, or a total shear time of 10 seconds, whichever is longer. We sample the rotation rate and torque with a rate of 10~Hz. The results for $h$ = 0.33 are shown in Fig.~\ref{fig:transient}, both the instantaneous speed of the disk and the driving torque as a function of time. Note the logarithmic scale on the time axis. We see that for $\Gamma = 0.2$, the set speeds are achieved very accurately and no large fluctuations or transients are observable. The fluctuations in the torque signal are however appreciable. Transients are short, even for the slowest run --- for more on transients, see the next section. For $\Gamma = 0.8$, the fluctuations in both speed and torque are strongly reduced and we can clearly see that no long time transients are present in these experiments. We conclude that the mean driving torque $T$ and mean rotation rate $\Omega$ are well defined quantities in steady state shear experiments.


\begin{thebibliography}{100}
\expandafter\ifx\csname natexlab\endcsname\relax\def\natexlab#1{#1}\fi
\expandafter\ifx\csname bibnamefont\endcsname\relax
  \def\bibnamefont#1{#1}\fi
\expandafter\ifx\csname bibfnamefont\endcsname\relax
  \def\bibfnamefont#1{#1}\fi
\expandafter\ifx\csname citenamefont\endcsname\relax
  \def\citenamefont#1{#1}\fi
\expandafter\ifx\csname url\endcsname\relax
  \def\url#1{\texttt{#1}}\fi
\expandafter\ifx\csname urlprefix\endcsname\relax\def\urlprefix{URL }\fi
\providecommand{\bibinfo}[2]{#2}
\providecommand{\eprint}[2][]{\url{#2}}

\bibitem[{\citenamefont{Duran}(1999)}]{1995_book_duran}
\bibinfo{author}{\bibfnamefont{J.}~\bibnamefont{Duran}},
  \emph{\bibinfo{title}{Sands, Powders and Grains: An Introduction to the
  Physics of Granular Materials}} (\bibinfo{year}{1999}).

\bibitem[{\citenamefont{Jaeger et~al.}(1996)\citenamefont{Jaeger, Nagel, and
  Behringer}}]{1996_revmodphys_jaeger}
\bibinfo{author}{\bibfnamefont{H.~M.} \bibnamefont{Jaeger}},
  \bibinfo{author}{\bibfnamefont{S.~R.} \bibnamefont{Nagel}}, \bibnamefont{and}
  \bibinfo{author}{\bibfnamefont{R.~P.} \bibnamefont{Behringer}},
  \bibinfo{journal}{Rev. Mod. Phys.} \textbf{\bibinfo{volume}{68}},
  \bibinfo{pages}{1259} (\bibinfo{year}{1996}).

\bibitem[{\citenamefont{MiDi}(2004)}]{2004_epje_gdrmidi}
\bibinfo{author}{\bibnamefont{GDR MiDi}}, \bibinfo{journal}{Eur.
  Phys. J. E} \textbf{\bibinfo{volume}{14}}, \bibinfo{pages}{341}
  (\bibinfo{year}{2004}).

\bibitem[{\citenamefont{Forterre and
  Pouliquen}(2008)}]{2008_annurevfluid_forterre}
\bibinfo{author}{\bibfnamefont{Y.}~\bibnamefont{Forterre}} \bibnamefont{and}
  \bibinfo{author}{\bibfnamefont{O.}~\bibnamefont{Pouliquen}},
  \bibinfo{journal}{Annu. Rev. Fluid Mech.} \textbf{\bibinfo{volume}{40}},
  \bibinfo{pages}{1} (\bibinfo{year}{2008}).

\bibitem[{\citenamefont{Nichol et~al.}(2010)\citenamefont{Nichol, Zanin,
  Bastien, Wandersman, and van Hecke}}]{2010_prl_nichol}
\bibinfo{author}{\bibfnamefont{K.}~\bibnamefont{Nichol}},
  \bibinfo{author}{\bibfnamefont{A.}~\bibnamefont{Zanin}},
  \bibinfo{author}{\bibfnamefont{R.}~\bibnamefont{Bastien}},
  \bibinfo{author}{\bibfnamefont{E.}~\bibnamefont{Wandersman}},
  \bibnamefont{and} \bibinfo{author}{\bibfnamefont{M.}~\bibnamefont{van
  Hecke}}, \bibinfo{journal}{Phys. Rev. Lett.} \textbf{\bibinfo{volume}{104}},
  \bibinfo{pages}{078302} (\bibinfo{year}{2010}).

\bibitem[{\citenamefont{S\'{a}nchez et~al.}(2007)\citenamefont{S\'{a}nchez,
  Raynaud, Lanuza, Andreotti, Cl\'{e}ment, and Aranson}}]{2007_pre_sanchez}
\bibinfo{author}{\bibfnamefont{I.}~\bibnamefont{S\'{a}nchez}},
  \bibinfo{author}{\bibfnamefont{F.}~\bibnamefont{Raynaud}},
  \bibinfo{author}{\bibfnamefont{J.}~\bibnamefont{Lanuza}},
  \bibinfo{author}{\bibfnamefont{B.}~\bibnamefont{Andreotti}},
  \bibinfo{author}{\bibfnamefont{E.}~\bibnamefont{Cl\'{e}ment}},
  \bibnamefont{and} \bibinfo{author}{\bibfnamefont{I.~S.}
  \bibnamefont{Aranson}}, \bibinfo{journal}{Phys. Rev. E}
  \textbf{\bibinfo{volume}{76}}, \bibinfo{eid}{060301} (\bibinfo{year}{2007}).

\bibitem[{\citenamefont{Rubin et~al.}(2006)\citenamefont{Rubin, Goldenson, and
  Voth}}]{2006_pre_rubin}
\bibinfo{author}{\bibfnamefont{D.}~\bibnamefont{Rubin}},
  \bibinfo{author}{\bibfnamefont{N.}~\bibnamefont{Goldenson}},
  \bibnamefont{and} \bibinfo{author}{\bibfnamefont{G.~A.} \bibnamefont{Voth}},
  \bibinfo{journal}{Phys. Rev. E} \textbf{\bibinfo{volume}{74}},
  \bibinfo{eid}{051307} (\bibinfo{year}{2006}).

\bibitem[{\citenamefont{Jaeger et~al.}(1989)\citenamefont{Jaeger, Liu, and
  Nagel}}]{1989_prl_jaeger}
\bibinfo{author}{\bibfnamefont{H.~M.} \bibnamefont{Jaeger}},
  \bibinfo{author}{\bibfnamefont{C.-h.} \bibnamefont{Liu}}, \bibnamefont{and}
  \bibinfo{author}{\bibfnamefont{S.~R.} \bibnamefont{Nagel}},
  \bibinfo{journal}{Phys. Rev. Lett.} \textbf{\bibinfo{volume}{62}},
  \bibinfo{pages}{40} (\bibinfo{year}{1989}).

\bibitem[{\citenamefont{Marchal et~al.}(2009)\citenamefont{Marchal, Smirani,
  and Choplin}}]{2009_jrheol_marchal}
\bibinfo{author}{\bibfnamefont{P.}~\bibnamefont{Marchal}},
  \bibinfo{author}{\bibfnamefont{N.}~\bibnamefont{Smirani}}, \bibnamefont{and}
  \bibinfo{author}{\bibfnamefont{L.}~\bibnamefont{Choplin}},
  \bibinfo{journal}{Journal of Rheology} \textbf{\bibinfo{volume}{53}},
  \bibinfo{pages}{1} (\bibinfo{year}{2009}).

\bibitem[{\citenamefont{Janda et~al.}(2009)\citenamefont{Janda, Maza,
  Garcimartin, Kolb, Lanuza, and Cl\'ement}}]{2009_epl_janda}
\bibinfo{author}{\bibfnamefont{A.}~\bibnamefont{Janda}},
  \bibinfo{author}{\bibfnamefont{D.}~\bibnamefont{Maza}},
  \bibinfo{author}{\bibfnamefont{A.}~\bibnamefont{Garcimartin}},
  \bibinfo{author}{\bibfnamefont{E.}~\bibnamefont{Kolb}},
  \bibinfo{author}{\bibfnamefont{J.}~\bibnamefont{Lanuza}}, \bibnamefont{and}
  \bibinfo{author}{\bibfnamefont{E.}~\bibnamefont{Cl\'ement}},
  \bibinfo{journal}{EPL} \textbf{\bibinfo{volume}{87}}, \bibinfo{pages}{24002}
  (\bibinfo{year}{2009}).

\bibitem[{\citenamefont{Reddy et~al.}(2011)\citenamefont{Reddy, Forterre, and
  Pouliquen}}]{2011_prl_reddy}
\bibinfo{author}{\bibfnamefont{K.~A.} \bibnamefont{Reddy}},
  \bibinfo{author}{\bibfnamefont{Y.}~\bibnamefont{Forterre}}, \bibnamefont{and}
  \bibinfo{author}{\bibfnamefont{O.}~\bibnamefont{Pouliquen}},
  \bibinfo{journal}{Phys. Rev. Lett.} \textbf{\bibinfo{volume}{106}},
  \bibinfo{pages}{108301} (\bibinfo{year}{2011}).

\bibitem[{\citenamefont{Nichol and van Hecke}(2012)}]{2012_pre_nichol}
\bibinfo{author}{\bibfnamefont{K.}~\bibnamefont{Nichol}} \bibnamefont{and}
  \bibinfo{author}{\bibfnamefont{M.}~\bibnamefont{van Hecke}},
  \bibinfo{journal}{Phys. Rev. E} \textbf{\bibinfo{volume}{85}},
  \bibinfo{pages}{061309} (\bibinfo{year}{2012}).

\bibitem[{\citenamefont{Wandersman and van Hecke}(2013)}]{2013_elie_arxiv}
\bibinfo{author}{\bibfnamefont{E.}~\bibnamefont{Wandersman}} \bibnamefont{and}
  \bibinfo{author}{\bibfnamefont{M.}~\bibnamefont{van Hecke}},
  \bibinfo{journal}{arXiv:1309.5238 [cond-mat.soft]}  (\bibinfo{year}{2013}).

\bibitem[{\citenamefont{Bouzid et~al.}(2013)\citenamefont{Bouzid, Trulsson,
  Claudin, Cl\'ement, and Andeotti}}]{andreotti_arxiv}  
\bibinfo{author}{\bibfnamefont{M.}~\bibnamefont{Bouzid}},
  \bibinfo{author}{\bibfnamefont{M.}~\bibnamefont{Trulsson}},
  \bibinfo{author}{\bibfnamefont{P.}~\bibnamefont{Claudin}},
  \bibinfo{author}{\bibfnamefont{E.}~\bibnamefont{Cl\'ement}}, \bibnamefont{and}
  \bibinfo{author}{\bibfnamefont{B.}~\bibnamefont{Andeotti}},
  \bibinfo{journal}{arXiv:1301.3308v3 [cond-mat.soft]}  (\bibinfo{year}{2013}).  

\bibitem[{\citenamefont{Falk and Langer}(1998)}]{1998_pre_falk}
\bibinfo{author}{\bibfnamefont{M.~L.} \bibnamefont{Falk}} \bibnamefont{and}
  \bibinfo{author}{\bibfnamefont{J.~S.} \bibnamefont{Langer}},
  \bibinfo{journal}{Phys. Rev. E} \textbf{\bibinfo{volume}{57}},
  \bibinfo{pages}{7192} (\bibinfo{year}{1998}).

\bibitem[{\citenamefont{Sollich et~al.}(1997)\citenamefont{Sollich, Lequeux,
  H\'ebraud, and Cates}}]{1997_prl_sollich}
\bibinfo{author}{\bibfnamefont{P.}~\bibnamefont{Sollich}},
  \bibinfo{author}{\bibfnamefont{F.}~\bibnamefont{Lequeux}},
  \bibinfo{author}{\bibfnamefont{P.}~\bibnamefont{H\'ebraud}},
  \bibnamefont{and} \bibinfo{author}{\bibfnamefont{M.}~\bibnamefont{Cates}},
  \bibinfo{journal}{Phys. Rev. Lett.} \textbf{\bibinfo{volume}{78}},
  \bibinfo{pages}{2020} (\bibinfo{year}{1997}).

\bibitem[{\citenamefont{Krimer et~al.}(2012)\citenamefont{Krimer, Mahle, and
  Liu}}]{2012_pre_krimer}
\bibinfo{author}{\bibfnamefont{D.~O.} \bibnamefont{Krimer}},
  \bibinfo{author}{\bibfnamefont{S.}~\bibnamefont{Mahle}}, \bibnamefont{and}
  \bibinfo{author}{\bibfnamefont{M.}~\bibnamefont{Liu}},
  \bibinfo{journal}{Phys. Rev. E} \textbf{\bibinfo{volume}{86}},
  \bibinfo{pages}{061312} (\bibinfo{year}{2012}).

\bibitem[{\citenamefont{Goyon et~al.}(2008)\citenamefont{Goyon, Colin, Ovarlez,
  Ajdari, and Bocquet}}]{2008_nature_goyon}
\bibinfo{author}{\bibfnamefont{J.}~\bibnamefont{Goyon}},
  \bibinfo{author}{\bibfnamefont{A.}~\bibnamefont{Colin}},
  \bibinfo{author}{\bibfnamefont{G.}~\bibnamefont{Ovarlez}},
  \bibinfo{author}{\bibfnamefont{A.}~\bibnamefont{Ajdari}}, \bibnamefont{and}
  \bibinfo{author}{\bibfnamefont{L.}~\bibnamefont{Bocquet}},
  \bibinfo{journal}{Nature} \textbf{\bibinfo{volume}{454}}, \bibinfo{pages}{84}
  (\bibinfo{year}{2008}).

\bibitem[{\citenamefont{Caballero-Robledo and
  Clément}(2009)}]{2009_epje_caballero}
\bibinfo{author}{\bibfnamefont{G.}~\bibnamefont{Caballero-Robledo}}
  \bibnamefont{and} \bibinfo{author}{\bibfnamefont{E.}~\bibnamefont{Cl\'ement}},
  \bibinfo{journal}{Eur. Phys. J. E Soft Matter} \textbf{\bibinfo{volume}{30}}, \bibinfo{pages}{395}
  (\bibinfo{year}{2009}).

\bibitem[{\citenamefont{Caballero et~al.}(2005)\citenamefont{Caballero, Kolb,
  Lindner, Lanuza, and Cl\'ement}}]{2005_jphyscm_caballero}
\bibinfo{author}{\bibfnamefont{G.}~\bibnamefont{Caballero}},
  \bibinfo{author}{\bibfnamefont{E.}~\bibnamefont{Kolb}},
  \bibinfo{author}{\bibfnamefont{A.}~\bibnamefont{Lindner}},
  \bibinfo{author}{\bibfnamefont{J.}~\bibnamefont{Lanuza}}, \bibnamefont{and}
  \bibinfo{author}{\bibfnamefont{E.}~\bibnamefont{Cl\'ement}},
  \bibinfo{journal}{J. Phys. Cond. Matt.} \textbf{\bibinfo{volume}{17}},
  \bibinfo{pages}{2503} (\bibinfo{year}{2005}).

\bibitem[{\citenamefont{Coulais et~al.}(2012)\citenamefont{Coulais, Behringer,
  and Dauchot}}]{2012_epl_coulais}
\bibinfo{author}{\bibfnamefont{C.}~\bibnamefont{Coulais}},
  \bibinfo{author}{\bibfnamefont{R.~P.} \bibnamefont{Behringer}},
  \bibnamefont{and} \bibinfo{author}{\bibfnamefont{O.}~\bibnamefont{Dauchot}},
  \bibinfo{journal}{EPL} \textbf{\bibinfo{volume}{100}},
  \bibinfo{pages}{44005} (\bibinfo{year}{2012}).

\bibitem[{\citenamefont{Longhi et~al.}(2002)\citenamefont{Longhi, Easwar, and
  Menon}}]{2002_prl_longhi}
\bibinfo{author}{\bibfnamefont{E.}~\bibnamefont{Longhi}},
  \bibinfo{author}{\bibfnamefont{N.}~\bibnamefont{Easwar}}, \bibnamefont{and}
  \bibinfo{author}{\bibfnamefont{N.}~\bibnamefont{Menon}},
  \bibinfo{journal}{Phys. Rev. Lett.} \textbf{\bibinfo{volume}{89}},
  \bibinfo{pages}{045501} (\bibinfo{year}{2002}).

\bibitem[{\citenamefont{Schall and van Hecke}(2010)}]{schall}
\bibinfo{author}{\bibfnamefont{P.}~\bibnamefont{Schall}} \bibnamefont{and}
  \bibinfo{author}{\bibfnamefont{M.}~\bibnamefont{van Hecke}},
  \bibinfo{journal}{Ann. Rev. Fluid Mech.} \textbf{\bibinfo{volume}{42}},
  \bibinfo{pages}{67} (\bibinfo{year}{2010}).

\bibitem[{\citenamefont{Crassous et~al.}(2008)\citenamefont{Crassous, Metayer,
  Richard, and Laroche}}]{2008_jstatmech_crassous}
\bibinfo{author}{\bibfnamefont{J.}~\bibnamefont{Crassous}},
  \bibinfo{author}{\bibfnamefont{J.-F.} \bibnamefont{Metayer}},
  \bibinfo{author}{\bibfnamefont{P.}~\bibnamefont{Richard}}, \bibnamefont{and}
  \bibinfo{author}{\bibfnamefont{C.}~\bibnamefont{Laroche}},
  \bibinfo{journal}{J. Stat. Mech.} p. \bibinfo{pages}{P03009}
  (\bibinfo{year}{2008}).

\bibitem[{\citenamefont{Amon et~al.}(2012)\citenamefont{Amon, Nguyen, Bruand,
  Crassous, and Cl\'ement}}]{2012_prl_amon}
\bibinfo{author}{\bibfnamefont{A.}~\bibnamefont{Amon}},
  \bibinfo{author}{\bibfnamefont{V.~B.} \bibnamefont{Nguyen}},
  \bibinfo{author}{\bibfnamefont{A.}~\bibnamefont{Bruand}},
  \bibinfo{author}{\bibfnamefont{J.}~\bibnamefont{Crassous}}, \bibnamefont{and}
  \bibinfo{author}{\bibfnamefont{E.}~\bibnamefont{Cl\'ement}},
  \bibinfo{journal}{Phys. Rev. Lett.} \textbf{\bibinfo{volume}{108}},
  \bibinfo{pages}{135502} (\bibinfo{year}{2012}).

\bibitem[{\citenamefont{Jagla}(2008)}]{2008_pre_jagla}
\bibinfo{author}{\bibfnamefont{E.~A.} \bibnamefont{Jagla}},
  \bibinfo{journal}{Phys. Rev. E} \textbf{\bibinfo{volume}{78}},
  \bibinfo{eid}{026105} (\bibinfo{year}{2008}).

\bibitem[{\citenamefont{T\"{o}r\"{o}k et~al.}(2007)\citenamefont{T\"{o}r\"{o}k,
  Unger, Kert\'{e}sz, and Wolf}}]{2007_pre_torok}
\bibinfo{author}{\bibfnamefont{J.}~\bibnamefont{T\"{o}r\"{o}k}},
  \bibinfo{author}{\bibfnamefont{T.}~\bibnamefont{Unger}},
  \bibinfo{author}{\bibfnamefont{J.}~\bibnamefont{Kert\'{e}sz}},
  \bibnamefont{and} \bibinfo{author}{\bibfnamefont{D.~E.} \bibnamefont{Wolf}},
  \bibinfo{journal}{Phys. Rev. E} \textbf{\bibinfo{volume}{75}},
  \bibinfo{eid}{011305} (\bibinfo{year}{2007}).

\bibitem[{\citenamefont{Depken et~al.}(2006)\citenamefont{Depken, van Saarloos,
  and van Hecke}}]{2006_pre_depken}
\bibinfo{author}{\bibfnamefont{M.}~\bibnamefont{Depken}},
  \bibinfo{author}{\bibfnamefont{W.}~\bibnamefont{van Saarloos}},
  \bibnamefont{and} \bibinfo{author}{\bibfnamefont{M.}~\bibnamefont{van
  Hecke}}, \bibinfo{journal}{Phys. Rev. E} \textbf{\bibinfo{volume}{73}},
  \bibinfo{eid}{031302} (\bibinfo{year}{2006}).

\bibitem[{\citenamefont{Kamrin and Koval}(2012)}]{2012_prl_kamrin}
\bibinfo{author}{\bibfnamefont{K.}~\bibnamefont{Kamrin}} \bibnamefont{and}
  \bibinfo{author}{\bibfnamefont{G.}~\bibnamefont{Koval}},
  \bibinfo{journal}{Phys. Rev. Lett.} \textbf{\bibinfo{volume}{108}},
  \bibinfo{pages}{178301} (\bibinfo{year}{2012}).

\bibitem[{\citenamefont{Henann and Kamrin}(2013)}]{PNAS_kamrin}
\bibinfo{author}{\bibfnamefont{D.~L.} \bibnamefont{Henann}} \bibnamefont{and}
  \bibinfo{author}{\bibfnamefont{K.}~\bibnamefont{Kamrin}},
  \bibinfo{journal}{PNAS} \textbf{\bibinfo{volume}{110}}, \bibinfo{pages}{6730}
  (\bibinfo{year}{2013}).

\bibitem[{\citenamefont{Dijksman et~al.}(2011)\citenamefont{Dijksman, Wortel,
  van Dellen, Dauchot, and van Hecke}}]{2011_prl_dijksman}
\bibinfo{author}{\bibfnamefont{J.~A.} \bibnamefont{Dijksman}},
  \bibinfo{author}{\bibfnamefont{G.~H.} \bibnamefont{Wortel}},
  \bibinfo{author}{\bibfnamefont{L.~T.~H.} \bibnamefont{van Dellen}},
  \bibinfo{author}{\bibfnamefont{O.}~\bibnamefont{Dauchot}}, \bibnamefont{and}
  \bibinfo{author}{\bibfnamefont{M.}~\bibnamefont{van Hecke}},
  \bibinfo{journal}{Phys. Rev. Lett.} \textbf{\bibinfo{volume}{107}},
  \bibinfo{pages}{108303} (\bibinfo{year}{2011}).

\bibitem[{\citenamefont{Nedderman and Laohakul}(1980)}]{1980_powtech_nedderman}
\bibinfo{author}{\bibfnamefont{R.~M.} \bibnamefont{Nedderman}}
  \bibnamefont{and} \bibinfo{author}{\bibfnamefont{C.}~\bibnamefont{Laohakul}},
  \bibinfo{journal}{Powd. Techn.} \textbf{\bibinfo{volume}{25}},
  \bibinfo{pages}{91 } (\bibinfo{year}{1980}).

\bibitem[{\citenamefont{Fenistein et~al.}(2004)\citenamefont{Fenistein, van~de
  Meent, and van Hecke}}]{2004_prl_fenistein}
\bibinfo{author}{\bibfnamefont{D.}~\bibnamefont{Fenistein}},
  \bibinfo{author}{\bibfnamefont{J.-W.} \bibnamefont{van~de Meent}},
  \bibnamefont{and} \bibinfo{author}{\bibfnamefont{M.}~\bibnamefont{van
  Hecke}}, \bibinfo{journal}{Phys. Rev. Lett.} \textbf{\bibinfo{volume}{92}},
  \bibinfo{eid}{094301} (\bibinfo{year}{2004}).

\bibitem[{\citenamefont{Fenistein et~al.}(2006)\citenamefont{Fenistein, van~de
  Meent, and van Hecke}}]{2006_prl_fenistein}
\bibinfo{author}{\bibfnamefont{D.}~\bibnamefont{Fenistein}},
  \bibinfo{author}{\bibfnamefont{J.-W.} \bibnamefont{van~de Meent}},
  \bibnamefont{and} \bibinfo{author}{\bibfnamefont{M.}~\bibnamefont{van
  Hecke}}, \bibinfo{journal}{Phys. Rev. Lett.} \textbf{\bibinfo{volume}{96}},
  \bibinfo{eid}{118001}  (\bibinfo{year}{2006}).

\bibitem[{\citenamefont{Dijksman and van Hecke}(2010)}]{2010_sm_dijksman}
\bibinfo{author}{\bibfnamefont{J.~A.} \bibnamefont{Dijksman}} \bibnamefont{and}
  \bibinfo{author}{\bibfnamefont{M.}~\bibnamefont{van Hecke}},
  \bibinfo{journal}{Soft Matter} \textbf{\bibinfo{volume}{6}},
  \bibinfo{pages}{2901} (\bibinfo{year}{2010}).

\bibitem[{\citenamefont{Cruz et~al.}(2002)\citenamefont{Cruz, Chevoir, Bonn,
  and Coussot}}]{coussot_pre_2002}
\bibinfo{author}{\bibfnamefont{D.~D.} \bibnamefont{Cruz}},
  \bibinfo{author}{\bibfnamefont{F.}~\bibnamefont{Chevoir}},
  \bibinfo{author}{\bibfnamefont{D.}~\bibnamefont{Bonn}}, \bibnamefont{and}
  \bibinfo{author}{\bibfnamefont{P.}~\bibnamefont{Coussot}},
  \bibinfo{journal}{Phys. Rev. E.} \textbf{\bibinfo{volume}{66}}
  (\bibinfo{year}{2002}).

\bibitem[{\citenamefont{M{\o}ller et~al.}(2009)\citenamefont{M{\o}ller, Fall, and
  Bonn}}]{bonn_epl_2009}
\bibinfo{author}{\bibfnamefont{P.}~\bibnamefont{M{\o}ller}},
  \bibinfo{author}{\bibfnamefont{A.}~\bibnamefont{Fall}}, \bibnamefont{and}
  \bibinfo{author}{\bibfnamefont{D.}~\bibnamefont{Bonn}},
  \bibinfo{journal}{Eur. Phys. Lett.} \textbf{\bibinfo{volume}{87}}
  (\bibinfo{year}{2009}).

\bibitem[{\citenamefont{Andrade}(1914)}]{andrade_1914}
\bibinfo{author}{\bibfnamefont{E.}~\bibnamefont{Andrade}},
  \bibinfo{journal}{Proc. R. Soc. A} \textbf{\bibinfo{volume}{90}}
  (\bibinfo{year}{1914}).

\bibitem[{\citenamefont{Petekidis et~al.}(2004)\citenamefont{Petekidis,
  Vlassopoulos, and Pusey}}]{2004_jphyscondmat_petekidis}
\bibinfo{author}{\bibfnamefont{G.}~\bibnamefont{Petekidis}},
  \bibinfo{author}{\bibfnamefont{D.}~\bibnamefont{Vlassopoulos}},
  \bibnamefont{and} \bibinfo{author}{\bibfnamefont{P.~N.} \bibnamefont{Pusey}},
  \bibinfo{journal}{J. Phys. Cond. Matt.} \textbf{\bibinfo{volume}{16}},
  \bibinfo{pages}{3955} (\bibinfo{year}{2004}).

\bibitem[{\citenamefont{Siebenb\"urger et~al.}(2012)\citenamefont{Siebenb\"urger,
  Ballauff, and Voightmann}}]{ballauff_prl_2012}
\bibinfo{author}{\bibfnamefont{M.}~\bibnamefont{Siebenb\"urger}},
  \bibinfo{author}{\bibfnamefont{M.}~\bibnamefont{Ballauff}}, \bibnamefont{and}
  \bibinfo{author}{\bibfnamefont{T.}~\bibnamefont{Voightmann}},
  \bibinfo{journal}{Phys. Rev. Lett.} \textbf{\bibinfo{volume}{108}}
  (\bibinfo{year}{2012}).

\bibitem[{\citenamefont{Unger et~al.}(2004)\citenamefont{Unger, T\"or\"ok,
  Kert\'esz, and Wolf}}]{2004_prl_unger}
\bibinfo{author}{\bibfnamefont{T.}~\bibnamefont{Unger}},
  \bibinfo{author}{\bibfnamefont{J.}~\bibnamefont{T\"or\"ok}},
  \bibinfo{author}{\bibfnamefont{J.}~\bibnamefont{Kert\'esz}},
  \bibnamefont{and} \bibinfo{author}{\bibfnamefont{D.~E.} \bibnamefont{Wolf}},
  \bibinfo{journal}{Phys. Rev. Lett.} \textbf{\bibinfo{volume}{92}},
  \bibinfo{pages}{214301} (\bibinfo{year}{2004}).

\bibitem[{\citenamefont{Wortel et~al.}(2014)}]{wortel_aniso}
\bibinfo{author}{\bibfnamefont{G.}~\bibnamefont{Wortel}}, \bibinfo{author}{\bibfnamefont{O.}~\bibnamefont{Dauchot}},
\bibnamefont{and} \bibinfo{author}{\bibfnamefont{M.}~\bibnamefont{van Hecke}}
  (\bibinfo{year}{2014}), \bibinfo{note}{in prep.}

\bibitem[{\citenamefont{Reynolds}(1885)}]{1885_philmag_reynolds}
\bibinfo{author}{\bibfnamefont{O.}~\bibnamefont{Reynolds}},
  \bibinfo{journal}{Phil. Mag.}  (\bibinfo{year}{1885}).

\bibitem[{\citenamefont{Pouliquen et~al.}(2006)\citenamefont{Pouliquen, Cassar,
  Jop, Forterre, and Nicolas}}]{pouliquen_I}
\bibinfo{author}{\bibfnamefont{O.}~\bibnamefont{Pouliquen}},
  \bibinfo{author}{\bibfnamefont{C.}~\bibnamefont{Cassar}},
  \bibinfo{author}{\bibfnamefont{P.}~\bibnamefont{Jop}},
  \bibinfo{author}{\bibfnamefont{Y.}~\bibnamefont{Forterre}}, \bibnamefont{and}
  \bibinfo{author}{\bibfnamefont{M.}~\bibnamefont{Nicolas}},
  \bibinfo{journal}{J. Stat. Mech.}  (\bibinfo{year}{2006}).

\bibitem[{\citenamefont{Jop et~al.}(2006)\citenamefont{Jop, Forterre, and
  Pouliquen}}]{2006_nature_jop}
\bibinfo{author}{\bibfnamefont{P.}~\bibnamefont{Jop}},
  \bibinfo{author}{\bibfnamefont{Y.}~\bibnamefont{Forterre}}, \bibnamefont{and}
  \bibinfo{author}{\bibfnamefont{O.}~\bibnamefont{Pouliquen}},
  \bibinfo{journal}{Nature} \textbf{\bibinfo{volume}{441}}, \bibinfo{eid}{727}
  (\bibinfo{year}{2006}).

\bibitem[{\citenamefont{Depken et~al.}(2007)\citenamefont{Depken, Lechman, van
  Hecke, van Saarloos, and Grest}}]{2007_epl_depken}
\bibinfo{author}{\bibfnamefont{M.}~\bibnamefont{Depken}},
  \bibinfo{author}{\bibfnamefont{J.~B.} \bibnamefont{Lechman}},
  \bibinfo{author}{\bibfnamefont{M.}~\bibnamefont{van Hecke}},
  \bibinfo{author}{\bibfnamefont{W.}~\bibnamefont{van Saarloos}},
  \bibnamefont{and} \bibinfo{author}{\bibfnamefont{G.~S.} \bibnamefont{Grest}},
  \bibinfo{journal}{EPL} \textbf{\bibinfo{volume}{78}}, \bibinfo{pages}{58001
  (5pp)} (\bibinfo{year}{2007}).

\bibitem[{\citenamefont{Fenistein and van Hecke}(2003)}]{2003_nature_fenistein}
\bibinfo{author}{\bibfnamefont{D.}~\bibnamefont{Fenistein}} \bibnamefont{and}
  \bibinfo{author}{\bibfnamefont{M.}~\bibnamefont{van Hecke}},
  \bibinfo{journal}{Nature} \textbf{\bibinfo{volume}{425}}, \bibinfo{eid}{256}
  (\bibinfo{year}{2003}).

\bibitem[{\citenamefont{Dijksman et~al.}(2010)\citenamefont{Dijksman,
  Wandersman, Slotterback, Berardi, Updegraff, van Hecke, and
  Losert}}]{2010_pre_dijksman}
\bibinfo{author}{\bibfnamefont{J.~A.} \bibnamefont{Dijksman}},
  \bibinfo{author}{\bibfnamefont{E.}~\bibnamefont{Wandersman}},
  \bibinfo{author}{\bibfnamefont{S.}~\bibnamefont{Slotterback}},
  \bibinfo{author}{\bibfnamefont{C.~R.} \bibnamefont{Berardi}},
  \bibinfo{author}{\bibfnamefont{W.~D.} \bibnamefont{Updegraff}},
  \bibinfo{author}{\bibfnamefont{M.}~\bibnamefont{van Hecke}},
  \bibnamefont{and} \bibinfo{author}{\bibfnamefont{W.}~\bibnamefont{Losert}},
  \bibinfo{journal}{Phys. Rev. E} \textbf{\bibinfo{volume}{82}},
  \bibinfo{pages}{060301} (\bibinfo{year}{2010}).

\bibitem[{\citenamefont{Wandersman et~al.}(2012)\citenamefont{Wandersman,
  Dijksman, and van Hecke}}]{wandesman_epl_2012}
\bibinfo{author}{\bibfnamefont{E.}~\bibnamefont{Wandersman}},
  \bibinfo{author}{\bibfnamefont{J.}~\bibnamefont{Dijksman}}, \bibnamefont{and}
  \bibinfo{author}{\bibfnamefont{M.}~\bibnamefont{van Hecke}},
  \bibinfo{journal}{Eur. Phys. Lett.}  (\bibinfo{year}{2012}).

\bibitem[{\citenamefont{Menon and Durian}(1997)}]{durian_prl_1997}
\bibinfo{author}{\bibfnamefont{N.}~\bibnamefont{Menon}} \bibnamefont{and}
  \bibinfo{author}{\bibfnamefont{D.}~\bibnamefont{Durian}},
  \bibinfo{journal}{Phys. Rev. Lett.} \textbf{\bibinfo{volume}{79}}
  (\bibinfo{year}{1997}).

\bibitem[{\citenamefont{M\"{o}bius et~al.}(2005)\citenamefont{M\"{o}bius,
  Cheng, Eshuis, Karczmar, Nagel, and Jaeger}}]{2005_pre_mobius}
\bibinfo{author}{\bibfnamefont{M.~E.} \bibnamefont{M\"{o}bius}},
  \bibinfo{author}{\bibfnamefont{X.}~\bibnamefont{Cheng}},
  \bibinfo{author}{\bibfnamefont{P.}~\bibnamefont{Eshuis}},
  \bibinfo{author}{\bibfnamefont{G.~S.} \bibnamefont{Karczmar}},
  \bibinfo{author}{\bibfnamefont{S.~R.} \bibnamefont{Nagel}}, \bibnamefont{and}
  \bibinfo{author}{\bibfnamefont{H.~M.} \bibnamefont{Jaeger}},
  \bibinfo{journal}{Phys. Rev. E} \textbf{\bibinfo{volume}{72}},
  \bibinfo{eid}{011304} (\bibinfo{year}{2005}).

\bibitem[{\citenamefont{Pak et~al.}(1995)\citenamefont{Pak, Van~Doorn, and
  Behringer}}]{1995_prl_pak}
\bibinfo{author}{\bibfnamefont{H.~K.} \bibnamefont{Pak}},
  \bibinfo{author}{\bibfnamefont{E.}~\bibnamefont{Van~Doorn}},
  \bibnamefont{and} \bibinfo{author}{\bibfnamefont{R.~P.}
  \bibnamefont{Behringer}}, \bibinfo{journal}{Phys. Rev. Lett.}
  \textbf{\bibinfo{volume}{74}}, \bibinfo{pages}{4643} (\bibinfo{year}{1995}).

\bibitem[{\citenamefont{Schr\"{o}ter et~al.}(2005)\citenamefont{Schr\"{o}ter,
  Goldman, and Swinney}}]{2005_pre_schroter}
\bibinfo{author}{\bibfnamefont{M.}~\bibnamefont{Schr\"{o}ter}},
  \bibinfo{author}{\bibfnamefont{D.~I.} \bibnamefont{Goldman}},
  \bibnamefont{and} \bibinfo{author}{\bibfnamefont{H.~L.}
  \bibnamefont{Swinney}}, \bibinfo{journal}{Phys. Rev. E}
  \textbf{\bibinfo{volume}{71}}, \bibinfo{eid}{030301} (\bibinfo{year}{2005}).

\end{thebibliography}
\end{document}